%% file: main.tex
\documentclass[fleqn,usenatbib,a4paper]{mnras}

\usepackage{newtxtext,newtxmath}

\usepackage[T1]{fontenc}
\usepackage{ae,aecompl}

\usepackage{graphicx}	
\usepackage{amsmath}	
\usepackage{amssymb}	
\usepackage{times}
\usepackage{tabularx}

\newcommand{\cn}{[\mathrm{C}/\mathrm{N}]}

\newcommand{\cm}{[\mathrm{C}/\mathrm{M}]}
\newcommand{\cfe}{[\mathrm{C}/\mathrm{Fe}]}

\newcommand{\nm}{[\mathrm{N}/\mathrm{M}]}

\newcommand{\feh}{[\mathrm{Fe}/\mathrm{H}]}
\newcommand{\mh}{[\mathrm{M}/\mathrm{H}]}

\newcommand{\om}{[\mathrm{O}/\mathrm{M}]}
\newcommand{\ofe}{[\mathrm{O}/\mathrm{Fe}]}
\newcommand{\afe}{[\alpha/\mathrm{Fe}]}

\newcommand{\am}{[\alpha/\mathrm{M}]}
\newcommand{\dex}{\,\mathrm{dex}}
\newcommand{\Gyr}{\,\mathrm{Gyr}}
\newcommand{\Myr}{\,\mathrm{Myr}}
\newcommand{\teff}{T_\mathrm{eff}}
\newcommand{\K}{\,\mathrm{K}}
\newcommand{\logg}{\log\,g}

\newcommand{\kpc}{\,\mathrm{kpc}}
\newcommand{\percent}{\,\mathrm{per\,cent}}
\newcommand{\bs}{\boldsymbol}
\newcommand{\angstrom}{\textup{\AA}}

\defcitealias{DasSanders2018}{DS18}

\title[$\sim3$ million ages from Gaia DR2]{Isochrone ages for $\sim3$ million stars with the second Gaia data release}

\author[J. L. Sanders and P. Das]{
Jason L. Sanders,$^{1}$\thanks{E-mail: jls@ast.cam.ac.uk (JLS)} 
Payel Das$^{2}$
\\
$^{1}$Institute of Astronomy, University of Cambridge, Madingley Road, Cambridge, CB3 0HA, UK\\
$^{2}$Rudolf Peierls Centre for Theoretical Physics, University of Oxford, OX1 3NP, UK
}

\date{Accepted XXX. Received YYY; in original form ZZZ}

\pubyear{2018}
\begin{document}
\label{firstpage}
\pagerange{\pageref{firstpage}--\pageref{lastpage}}
\maketitle

\begin{abstract}
We present a catalogue of distances, masses and ages for $\sim3$ million stars in the second Gaia data release with spectroscopic parameters available from the large spectroscopic surveys: APOGEE, Gaia-ESO, GALAH, LAMOST, RAVE and SEGUE. We use a Bayesian framework to characterise the probability density functions of distance, mass and age using photometric, spectroscopic and astrometric information, supplemented with spectroscopic masses where available for giant stars. Furthermore, we provide posterior extinction estimates ($A_V$) to every star using published extinction maps as a prior input. We provide an appendix with extinction coefficients for Gaia photometry derived from stellar models, which account for variation with intrinsic colour and total extinction. Our pipeline provides output estimates of the spectroscopic parameters, which can be used to inform improved spectroscopic analysis. We complement our catalogues with Galactocentric coordinates and actions with associated uncertainties. As a demonstration of the power of our catalogue, we produce velocity dispersion profiles of the disc separated by age and Galactocentric radius (between $3$ and $15\kpc$ from the Galactic centre). This suggests that the velocity dispersion profiles flatten with radius in the outer Galaxy ($>8\kpc$) and that at all radii the velocity dispersion follows the smooth power law with age observed in the solar neighbourhood.
\end{abstract}




\section{Introduction}
The second Gaia data release \citep[Gaia DR2][]{GaiaDR2} heralds a revolution in the study of the Milky Way. Gaia \citep{Gaia2016} has provided parallaxes and proper motions for $\sim1.3$ billion Milky Way stars \citep{GaiaDR2Astrometry} as well as highly accurate multi-band photometry for many of these stars \citep{GaiaDR2Photometry1,GaiaDR2Photometry2}. For a subset of stars brighter than $12$th magnitude in the narrow Gaia Radial Velocity Spectrometer (RVS) photometric band ($G_\mathrm{RVS}$), radial velocities have been computed from the Gaia RVS spectra \citep{GaiaDR2RVS1,GaiaDR2RVS2}. Whilst in its own right this dataset will lead to rapid advances in our knowledge of the structure of the Milky Way \citep{GaiaMWKinematics}, when complemented with other photometric surveys and large-scale spectroscopic surveys, Gaia is the perfect tool for Galactic archaeology studies.

The primary goal of Galactic archaeology is to construct a census of the structure of the Milky Way in spatial, kinematic and chemical space \citep{BlandHawthornGerhard2016}. Differences in the 
chemodynamical structure of populations reflects both differences between the chemodynamical structure of the populations at their formation epochs as well as differences in their subsequent dynamical evolution. The secondary goal of Galactic archaeology is to use the census of the Galaxy to infer the properties of the
Galaxy
at different epochs of star formation and hence the evolution history of the Galaxy from this data. This must necessarily be informed by Milky-Way-like simulations and models. To disentangle the impacts of dynamical evolution from the formation environment in this way, we require the rich, detailed extended picture of our Galaxy, provided by the synergy of Gaia with large-scale spectroscopic surveys. One question of interest for example is to understand whether the correlation between stellar age and velocity dispersion is a consequence of older stars being born hotter, or stars heating over time.

Before Gaia, reliable kinematics (from parallaxes) were only available in the solar neighbourhood \citep{Holmberg2009,Bensby2014}. Early extended studies of the Milky Way were limited to using large-scale photometric surveys \citep[e.g. 2MASS, SDSS][]{2MASS, SDSSDR14}. In recent years, large-scale spectroscopic surveys have been the dominant tool in furthering the study of the Milky Way.
Many spectroscopic surveys \citep[APOGEE, LAMOST, RAVE, GALAH, Gaia-ESO, SEGUE][]{APOGEE,Deng2012,Steinmetz2006,DeSilva2015,Gilmore2012,Yanny2009} have been designed with the intention of complementing the astrometry from Gaia with radial velocity information as well as detailed chemical abundances. With full phase-space coordinates, the dynamical structure of the Galaxy well beyond the solar neighbourhood can now be mapped in detail. Going forward, future surveys such as WEAVE \citep{WEAVE}, 4-MOST \cite{4MOST}, Milky Way Mapper \citep{SDSSV} and MOONS \citep{MOONS} will further complement the Gaia data and provide a more complete picture of the populations of the Milky Way.

The combination of spectroscopy, photometry and astrometry allows accurate characterization of stellar properties. Given a set of stellar models, this data combination can give accurate measurements of the distance, mass, age and other spectroscopic properties of the stars. In time, the Gaia data will be used to produce improvements in stellar models. However, a first step is to inspect the properties of the stars given the currently available models. Measuring distances from photometry and spectroscopy via sets of stellar isochrones has been developed over a number of years \citep{PontEyer2004,JorgensenLindegren2005,BurnettBinney,Binney2014} and has been widely applied to current spectroscopic surveys \citep[e.g.][]{Queiroz2018, Mints2018}. Utilising additional information from parallaxes \citep[e.g.][]{deSilva2006,McMillan2017} and/or masses \citep{DasSanders2018} is a natural extension. We highlight two synergies of the methodology.
For a typical star with $G<15$, Gaia DR2 provides a median parallax uncertainty of $\sim0.04\,\mathrm{mas}$ giving parallaxes precise to $20\percent$ out to $\sim5\kpc$. For more distant giants observed by spectroscopic surveys, spectro-photometric distances can be superior. The combination of spectro-photometry and astrometric information is particularly powerful \citep{McMillan2017} -- for instance, Gaia parallaxes can cleanly distinguish between nearby dwarfs and distant giants, producing improved spectroscopic parameter estimates. These estimates are useful initial guesses for improved analyses of the spectra.

The real advantage, however, of comparing astrometry and spectro-photometry to sets of stellar models is estimating stellar ages. The age of a star is a fundamental parameter in understanding the formation and evolution of the Galaxy, but, except in a few limited cases, can only be inferred via models \citep{Soderblom2010}. The most reliable age estimates are estimated for (near) co-eval cluster stars. Ages of co-eval populations are primarily constrained by the location of the main-sequence turn-off \citep[e.g.][]{MarinFranch2009} as a star's initial mass determines the time at which it leaves the main sequence. Such an approach can be utilised in the single field star case for those stars that are transitioning from the main sequence to the giant branch \citep[e.g.][]{Xiang2015,Xiang2017}. However, even with parallaxes from Gaia there is a strong degeneracy between metallicity and age that must be broken with the use of spectroscopic metallicities \citep{Howes2018}. Other populations of stars (e.g. lower main sequence dwarfs and giants) are much more difficult to accurately date due to only subtle differences in their observed spectro-photometric properties. However, recent theoretical and empirical work \citep{Masseron2015,Martig2016} has demonstrated that there are clear spectroscopic indicators of a giant star's mass (carbon and nitrogen atomic and molecular lines) and hence age \citep[e.g.][]{Ho2017,Wu2018}. When further combined with information on a giant's luminosity from Gaia parallaxes, it is also possible to accurately date giant stars. \cite{DasSanders2018} have recently demonstrated the power of including masses for giant stars in the standard isochrone pipeline.


In this paper, we provide distances, masses, ages, extinctions and spectroscopic parameters ($\teff$, $\logg$ and $\mh$) for stars from spectroscopic surveys combined with the Gaia data. In Section~\ref{Sec::Method} we describe the method employed, focussing on details of the extinction priors and extinction law, the scheme for estimating masses from spectroscopic parameters (`spectroscopic mass estimates') and the adopted Milky Way prior. In Section~\ref{Sec::Datasets}, we describe the datasets to which we have applied our algorithm, along with the Gaia data employed. In Section~\ref{Sec::Results}, we discuss the results of our procedure concentrating on the quality of the catalogue as well as indicating the possible chemo-dynamical studies that such a catalogue makes possible. We present our conclusions in Section~\ref{Sec::Conclusions}.

\section{Method}\label{Sec::Method}
We lay out the framework used to construct a probability density function for a star's distance. We assume the stars are single \citep[see][for handling binary stars]{Coronado2018} and are solely the products of single star evolution (which does not encompass the significant blue straggler population). We will see that assuming single stars can lead to identification of potential multiple systems a posteriori. We first present the framework using the isochrones in this subsection, discuss extinction in Section \ref{Sec::Extinction}, discuss complementing our data with spectroscopic mass estimates in Section \ref{Sec::Mass}, present our choice of Galaxy prior in Section \ref{Sec::Prior} and discuss the outputs of our approach in Section \ref{Sec::Outputs}.

We use the Bayesian method presented in \cite{BurnettBinney}. As discussed in \cite{BailerJones2018} and \cite{Luri2018}, when working with Gaia parallaxes it is necessary to employ a full Bayesian scheme including a reasonable distance prior to yield meaningful results (particularly in the cases of parallax measurements of order the uncertainty or negative parallaxes). We assess the probability of the $i$th star being at a distance $s$ (with distance modulus $\mu=10+5\log_{10}[s/\mathrm{kpc}]$) given the data $D_i$, which consists of observed spectroscopic data $S_i=(\logg, \teff, \mh)_i$, photometric data $m_i=(J, H,K_s,\cdots)_i$, astrometric data (Galactic coordinates $\ell_i,b_i$ and parallax $\varpi_i$) and in some cases mass estimates ($\mathcal{M}_i$ from spectroscopic mass estimators calibrated with asteroseismology) with associated uncertainties. The uncertainties in the spectroscopic parameters are given by the spectroscopic parameter covariance matrix $\Sigma$. The uncertainties in the photometry are assumed to be uncorrelated and given by $\sigma_{m,i}$, whilst the parallax and mass have associated uncertainties $\sigma_{\varpi,i}$ and $\sigma_{\mathcal{M},i}$ respectively. We drop the subscript $i$ in the following, for clarity. The resulting pdf is given by
\begin{equation}
p(s|D) = \frac{p(D|s)p(s)}{p(D)} \propto \int \mathrm{d}I\,p(D|I,s) p(s|I) p(I),
\label{Bayes}
\end{equation}
where the integral is performed over a set of isochrones $\{I\}$ indexed by metallicity, $\mh$, and log-age, $\log_{10}\tau$. We work with a set of PARSEC isochrones \citep[v1.2S -- excludes thermally pulsing AGB phase and using Reimers mass loss $\eta=0.2$,][]{Bressan2012,Chen2014,Tang2014,Chen2015} spaced by $0.01\dex$ in $\mh$ and $0.05\dex$ in $\log_{10}\tau$ (up to a maximum age of $12.6\Gyr$). Each isochrone gives a set of observed properties as a function of the initial mass $\mathcal{M}_0$. The integral over isochrones can therefore be written as
\begin{equation}
\int \mathrm{d}I =\int \mathrm{d}\mh\,\mathrm{d}\tau\,\mathrm{d}\mathcal{M}_0 = \sum_I \Delta \mh\Delta\tau\Delta \mathcal{M}_0,
\label{IsoDiscrete}
\end{equation}
where $\Delta \mh\,\Delta\tau\Delta \mathcal{M}_0$ is the volume occupied by each isochrone point $I$. Given an isochrone point and a distance $s$, the model spectroscopic $S'=(\logg, \teff, \mh$), photometric (absolute magnitudes $M'$) properties and current mass ($\mathcal{M}'$) may be computed and compared with the observed properties given the reported errors. We write the likelihood term
\begin{equation}
p(D|I,s) = p(m|M',s)p(S|S')p(\varpi|s)p(\mathcal{M}|S').
\end{equation}
and expand the first term as
\begin{equation}
\begin{split}
p(m|M',s) &= \int\mathrm{d}\ln A_V\, p(m|M',s,\ln A_V) p(\ln A_V), \\
p(\ln A_V) &= \mathcal{N}(\ln A_V-\langle\ln A_V\rangle,\sigma_{\ln A_V}),\\
\end{split}
\end{equation}
to marginalize over the unknown $V$-band extinction $A_V$. 
The four likelihood terms are then given by
\begin{equation}
\begin{split}
p(m|M',s,A_V)&=\prod_k \mathcal{N}(m_k-M'_k-\mu(s)-A_k,\sigma_{mk}),\\
p(S|S') &= \frac{1}{\sqrt{(2\pi)^3|\Sigma|}}\exp\Big(-\frac{1}{2} (S-S')\Sigma^{-1}(S-S')\Big),\\
p(\varpi|s)&=\mathcal{N}(\varpi-1/s,\sigma_{\varpi}),\\
p(\mathcal{M}|S')&=\mathcal{N}(\mathcal{M}-\mathcal{M}',\sigma_{\mathcal{M}}),
\end{split}
\end{equation}
where we have introduced the notation $\mathcal{N}(\mu,\sigma)$ for a Gaussian with mean $\mu$ and standard deviation $\sigma$, and $A_k$ is the extinction in the $k$th photometric band. The other term in our posterior is the prior $p(s|I) p(I)$, which we detail in Section~\ref{Sec::Prior}.

\subsection{Extinction}\label{Sec::Extinction}

When using photometry in our pipeline, we have to correct for the line-of-sight extinction. We choose to characterise this by the parameter $A_V$, the extinction in the Johnson $V$ band. The total extinction $A_V$ is related to the selective extinction $E(B-V)=A_B-A_V$ by $R(V)=A_V/E(B-V)$. Therefore, the extinction coefficient for the $V$ band, $R(V)$, is dependent on the assumed extinction law.
For photometric bands other than $V$ the extinction is given by $A_i=R(i)E(B-V)$ so is simply related to our parameter $A_V$ via $A_i = A_V R(i)/R(V)$. The set of coefficients $R(i)/R(V)$ describe the adopted extinction law whilst $A_V$ scales the total extinction. We first describe our choice of $R(i)/R(V)$ before discussing priors on $A_V$.

\subsubsection{Extinction coefficients}
Here we briefly describe the adopted extinction coefficients, $R(i)$. Full details are given in Appendix~\ref{Appendix::Extinction}. 
We use the extinction curve $A(\lambda)$ (total extinction at wavelength $\lambda$) from \cite{Schlafly2016} calibrated using APOGEE data. We scale $A(\lambda)$ to the units $E'$ adopted in the extinction maps of \citet[][$E'$ is slightly different from $E(B-V)$]{Green2017}. Therefore, the provided coefficients $R(i)$ can be used in combination with the \cite{Green2017} extinction map to find the extinction in band $i$ as $A_i=R(i)E'$.

Using the stellar model with $\teff=4500\K$ and $\logg=4.5$ from \cite{CastelliKurucz2004}, we compute $R(i)/R(V)$ for all photometric bands of interest (SDSS $u,g,r,i,z$, Pan-STARRS $g_\mathrm{P},r_\mathrm{P},i_\mathrm{P},z_\mathrm{P}$, 2MASS $J,H,K_s$, APASS $B, V$, WISE $W_1, W_2$, Gaia $G, G_\mathrm{BP}, G_\mathrm{RP}, G_\mathrm{RVS}$) by integrating the photometric band response over the spectrum.
For the broad Gaia $G$ band we consider variation of $R(G)$ with effective temperature by tabulating $R(G)$ over the full range of \cite{CastelliKurucz2004} models. In Appendix~\ref{Appendix::Extinction}, we provide expressions for $R(G)$ as a function of effective temperature and intrinsic colours in different bands.

There is also variation of $R(G)$ with $A_V$. We evaluate equation~\eqref{eqn::extinction} at a range of monochromatic extinctions and measure the gradient $\mathrm{d}R(i)/\mathrm{d}A_V$. The results are provided in Appendix~\ref{Appendix::Extinction}. For computational reasons, we do not consider variation of the extinction law (characterised by the \emph{parameter} $R_V$) despite \cite{Schlafly2016} demonstrating that there is variation of $\sim0.2$ across the APOGEE fields.

\subsubsection{Extinction prior}
Our prior $p(\ln A_V)$ uses extinction measurements from a combination of three extinction maps \citep[c.f.][]{Bovy2016}. We preferentially use the extinction maps from \citet[using the \texttt{dustmaps} \texttt{bayestar} interface]{Green2017}. For each star that falls in the Pan-STARRS footprint, we draw $10$ samples of the extinction $E'$ (in the \texttt{bayestar} 2017 units) at a set of discrete distances along the line-of-sight. At each distance, we use the coefficients from the previous section to compute the mean $\langle\ln A_V\rangle$ and the standard deviation $\sigma_{\ln A_V}$. 
For any star that falls outside the Pan-STARRS footprint (which only extends to $\delta\sim-30\deg$), we next attempt to use the extinction map from \cite{Marshall2006} which is confined to $|l|<100\deg$ and $|b|<10\deg$ and provides mean $K_s$ extinction and its uncertainty. For each star, we use $R(V)/R(K_s)$ from the previous section to find $\langle\ln A_V\rangle$ and $\sigma_{\ln A_V}$ on a grid in distance. Finally, where neither of these maps are available we use the 3D extinction map (expressed in $A_V$) from \cite{Drimmel2003} using the interface from \cite{Bovy2016}. Again we tabulate $\langle\ln A_V\rangle$ on a grid in distance and assume $30\percent$ uncertainty in $A_V$.

\subsection{Mass estimates}\label{Sec::Mass}
For giant stars, a mass measurement is a near direct measurement of age. Asteroseismology provides an estimate of stellar mass given measurements of the frequency spectrum of a star's oscillations. However, such observations require high quality photometry over a long time baseline so are limited to small subsets of stars confined to limited regions of the Galaxy (e.g. the Kepler field). Recent work \citep{Masseron2015,Martig2016,Ness2016} has demonstrated empirically that the [C/N] ratio in giant stars is an indicator of mass. Theoretical expectation \citep{Charbonnel1994} is that both the equilibrium position of core CNO burning and strength of dredge-up are functions of stellar mass. Higher mass stars produce more nitrogen than carbon leading to a suppression in [C/N]. We wish to use the carbon and nitrogen abundances as constraints within our framework by converting these spectroscopic parameters into estimates of the stellar mass.

We adopt the procedure in \citet[DS18]{DasSanders2018} for relating spectroscopic parameters to mass. This involves constructing a Bayesian artificial neural network for the input spectroscopic parameters $X=(\teff, \logg, \mh, \am, \cm, \nm)$ 
and their uncertainties $\sigma_X$ to
the mass $\mathcal{M}$ and its uncertainty $\sigma_\mathcal{M}$ (\citetalias{DasSanders2018} use the procedure to estimate mass, age and distance). We scale both the input $X$ and output parameters $\mathcal{M}$ to approximate unit gaussians. The neural network architecture presented here differs slightly with that in \citetalias{DasSanders2018}. Here the neural network contains two hidden layers (with 24 hidden nodes each) rather than one. Although this is a more complex architecture, marginalizing over the model parameters does not result in significant over-fitting. We again assume each layer (except the output) uses a sigmoid function $\sigma$
\begin{equation}
f_\theta(X)=w_o\sigma(w_2\sigma(w_1X+b_1)+b_2)+b_o.
\end{equation}
The weights $w_i$ are matrices and the biases $b_i$ vectors. To find the posterior distributions on the neural network parameters $\theta=(b_i,w_i)$, we evaluate
\begin{equation}
p(\theta|X,\mathcal{M})=p(X,\mathcal{M}|\theta)p(\theta).
\end{equation}
The likelihood term is given by
\begin{equation}
p(X,\mathcal{M}|\theta)=\int\mathrm{d}X'\,\mathcal{N}(X-X',\sigma_X)\mathcal{N}(\mathcal{M}-f_\theta(X'),\sigma_\mathcal{M}),
\end{equation}
and we choose normal priors on $w_i$ with zero mean and standard deviations given by hyperparameters $r$ -- one for each of the six parameters $w_1,w_2,w_o,b_1,b_2,b_o$. The inclusion of these hyperparameters is a development from \citetalias{DasSanders2018} and are important as, although the inputs and outputs have been scaled to approximate normal distributions, the hidden layers can have considerably larger dynamic range.

We implement this model in PyMC3 \citep{Salvatier2016} and train using automatic differentiation variational inference (ADVI). As our training set we use the overlap of APOGEE DR14 \citep[][described later]{SDSSDR14} with asteroseismic results \citep{Pinsonneault2014,Vrard2016}. We preferentially work with the \cite{Vrard2016} sample and supplement with \cite{Pinsonneault2014} APOKASC results for those stars not in \cite{Vrard2016}. We remove stars with \texttt{ASPCAPFLAG} $\neq0$, 
those without all elements of $X$ measured and duplicates. We further remove those stars that have been identified as rotating by \cite{Tayar2015}.

For unseen data, the posterior distribution on the mass is given by
\begin{equation}
p(\mathcal{M}|X)=\int \mathrm{d}X'\,\mathrm{d}\theta\,p(\mathcal{M}|X',\theta)\mathcal{N}(X-X',\sigma_X)p(\theta),
\end{equation}
where the first term is a $\delta$-function involving $f_\theta(X')$. We generate samples from this pdf by using the trace output from PyMC3 for the model parameters $\theta$ and sample from a Gaussian for the spectroscopic parameters. We reduce the resulting pdf to the first two moments: mean and standard deviation. We only apply this method if the inputs satisfy $4000<T_\mathrm{eff}/\,\mathrm{K}<5250$, $1<\log g<3.3$ and $-1.5<\mh<0.5$, which is the approximate parameter range covered by the training set.

Here we have used ADVI to characterise the pdf whereas \citetalias{DasSanders2018} fully sampled from the pdf using the No-U-Turn sampler \citep[NUTS,][]{NUTS}. The disadvantage of ADVI is that we fail to characterise the (potentially significant) correlations and multi-modality of the pdf (although highly isolated modes will not be well sampled by NUTS). Additionally, \citetalias{DasSanders2018} found the behaviour of NUTS to be smoother when varying the architecture. However, ADVI is considerably faster and in practice the results obtained are quite similar. 

The advantage of employing this method is that we need not perform quality cuts on our samples (provided we trust the reported uncertainties), and the uncertainty in the mass estimate reflects both the input uncertainty and the model uncertainty. Modelling the training set permits some level of de-noising of the output masses. The output mass uncertainty for unseen data is comparable (or better than) the mass uncertainty in the training set, significantly extending the power of asteroseismology to vast numbers of stars. Additionally, the flexibility of the model allows us to extrapolate into regions of spectroscopic parameter space that are sparsely populated where our model becomes more uncertain. A disadvantage is we are utilising spectroscopic information e.g. $\teff$ to constrain the mass and both mass and $\teff$ are then used in the distance pipeline. We are therefore using some spectroscopic information twice. However, using solely $\teff$, $\logg$ and $\mh$ produces poor constraints on the mass, and most of the constraint on the mass comes from $\cm$ and $\nm$ which are not further utilised. The use of the other spectroscopic parameters can be thought of as weakly adjusting the relationship between $\cm$ and $\nm$, and mass. Furthermore, we must assume that the stars within our asteroseismic sample are representative of the stars in our entire spectroscopic sample. Finally, we are required to use input spectroscopic parameters calibrated on the same scale as the training set.

\subsection{Galaxy prior}\label{Sec::Prior}
We choose to adopt a non-uniform prior distribution in distance, age and metallicity reflecting the prior knowledge that the considered stars belong primarily to the Galactic disc. For this we introduce the 3D prior distribution $p_\mathrm{gal}(\boldsymbol{x},\mh,\tau)$ with which we must include a Jacobian factor $s^2$ to relate to the prior on distance. Additionally, we use the Kroupa initial mass function \citep{Kroupa} $p(\mathcal{M}_0)$ as a prior on the initial mass:
\begin{equation}
p(s|I)p(I) = s^2f_\mathrm{prior}(\boldsymbol{x},\mh,\tau,\mathcal{M}_0)=s^2p_\mathrm{gal}(\boldsymbol{x},\mh,\tau)p(\mathcal{M}_0).
\label{Eqn::IsochronePrior}
\end{equation}
We decompose our Galaxy prior into multiple components
\begin{equation}
p_\mathrm{gal}(\boldsymbol{x},\mh,\tau)=\sum_ip_{\mathrm{gal},i}(\boldsymbol{x},\mh,\tau).
\end{equation}
Following \cite{Binney2014}, we use a three-component (thin disc, thick disc and halo prior) but supplemented by a bulge prior \citep[c.f.][]{Queiroz2018}. As we are interested in producing reliable age estimates, we follow \cite{Queiroz2018} and adopt a smooth age prior (as opposed to the truncated age prior from \cite{Binney2014} that e.g. assigns zero probability to young ages for high latitude distant stars). Each component can be written in the separable form
\begin{equation}
p_\mathrm{gal,i}(\boldsymbol{x},\mh,\tau)=p_\mathrm{gal,i}(\boldsymbol{x})p_\mathrm{gal,i}(\mh)p_\mathrm{gal,i}(\tau).
\end{equation}
We detail each component in turn. All age distributions are truncated at $12.6\,\mathrm{Gyr}$ (the largest isochrone age we consider) and normalized appropriately.
\begin{enumerate}
\item Thin disc: Double exponential profile (exponential in both $R$ and $z$) with scalelength $R_\mathrm{d}=2.6\,\mathrm{kpc}$ and scaleheight $z_\mathrm{d}=0.3\,\mathrm{kpc}$ normalized with local density $0.04\,M_\odot\,\mathrm{pc}^{-3}$ \citep{Bovy2018}, gaussian distribution in metallicity (mean $-0.1\,\mathrm{dex}$, standard deviation $0.3\,\mathrm{dex}$), age distribution
\begin{equation}
p_\mathrm{gal,thin}(\tau) = \begin{cases}
\mathrm{exp}(\tau/(8.4\,\mathrm{Gyr}))&\tau<8\,\mathrm{Gyr},\\2.6\mathrm{exp}(-\tfrac{1}{2}(\tau-8\,\mathrm{Gyr})^2/(1.5\,\mathrm{Gyr})^2)&\tau>8\,\mathrm{Gyr}.
\end{cases}
\end{equation}
\item Thick disc: Double exponential profile with scalelength $R_\mathrm{d}=2.0\,\mathrm{kpc}$ and scaleheight $z_\mathrm{d}=0.9\,\mathrm{kpc}$ normalized with local density $0.04\times0.04\,M_\odot\,\mathrm{pc}^{-3}$ \citep{BlandHawthornGerhard2016,Bovy2018}, gaussian distribution in metallicity (mean $-0.6\,\mathrm{dex}$, standard deviation $0.5\,\mathrm{dex}$), truncated gaussian in age (mean $10\,\mathrm{Gyr}$, standard deviation $2\,\mathrm{Gyr}$)
\item Halo: spherical power-law with $r^{-3.39}$ \citep{Binney2014} normalized with local density $0.005\times0.04\,M_\odot\,\mathrm{pc}^{-3}$ \citep{BlandHawthornGerhard2016,Bovy2018}, gaussian in metallicity (mean $-1.6\,\mathrm{dex}$, standard deviation $0.5\,\mathrm{dex}$), truncated gaussian in age (mean $12\,\mathrm{Gyr}$, standard deviation $2\,\mathrm{Gyr}$),
\item Bulge: Besan\c con density profile \citep{Robin2012}:
\begin{equation}
p_\mathrm{gal,bulge}(\boldsymbol{x}) = \mathrm{sech}^2\Big\{\Big[\Big(\frac{x'}{x_b}\Big)^{c_\perp}+\Big(\frac{y'}{y_b}\Big)^{c_\perp}\Big]^{c_{||}/c_\perp}+\Big(\frac{z}{z_b}\Big)^c_{||}\Big\}f(R),
\end{equation}
where $f(R)=\mathrm{e}^{-(R-R_c)^2/R_s^2}$ if $R>R_c$. $R$ is cylindrical polar radius and primed coordinates are aligned with the bar \citep[at an angle $19.57\,\mathrm{deg}$ relative to $l=0$,][]{Simion2017}. We
normalize the profile such that the central density is $35.45\,M_\odot\,\mathrm{pc}^{-3}$ \citep{Robin2012}, and we set $R_c=2.54\,\mathrm{kpc}$ and $R_s=0.5\,\mathrm{kpc}$ from \cite{Sharma2011} and all other parameters from \cite{Simion2017} S model fits to VVV data, gaussian in metallicity (mean $0\,\mathrm{dex}$, standard deviation $0.5\,\mathrm{dex}$), truncated gaussian in age (mean $10\,\mathrm{Gyr}$, standard deviation $3\,\mathrm{Gyr}$).
\end{enumerate}

Importantly, we place the Sun at radius $R_0=8.2\,\mathrm{kpc}$ and height above the plane $z_0=15\,\mathrm{pc}$ \citep{BlandHawthornGerhard2016}.

\subsection{Outputs}\label{Sec::Outputs}
With our specification, the full pdf of distance can be constructed using equation~\eqref{Bayes}. In practice, we simply compute the first two moments of the distance distribution 
e.g.
\begin{equation}
\langle \mu \rangle = \Big(\int \mathrm{d}I\,\mathrm{d}s\,\mu\,p(D|I,s) p(I,s)\Big) \Big/ \Big(\int \mathrm{d}I\mathrm{d}s\,p(D|I,s) p(I,s)\Big).
\label{Eq::moments}
\end{equation}
For this we sum over a range in distance modulus at each isochrone point given by $m-M-A_m\pm N\sigma_{m}$ where $\sigma_{m}$ is the uncertainty in the measured apparent magnitude $m$ and $M$ is the absolute magnitude of the isochrone point (we use $m=J$ or $m=J_\mathrm{VISTA}$). For speed, the extinction prior $p(\ln A_V)$ is only computed at the central value $m-M-A_m$. Additionally, the sum in equation~\eqref{IsoDiscrete} is only performed over the isochrone points within $N$ standard deviations of the observed metallicity and then we only consider points within $N$ standard deviations of the reported $\log g$ and $T_\mathrm{eff}$. We choose $N=5$ but increase this (iteratively by a factor of two) if there is no reported overlap with any isochrone points on a first pass. The range of integration for $\ln A_V$ is found from $\langle \ln A_V\rangle(s_0)\pm3\sigma_{\ln A_V}(s_0)$ evaluated at a first estimate of the distance $s_0$ neglecting extinction (using only infra-red photometry).

In a similar fashion to equation~\eqref{Eq::moments}, we can compute the moments of the log-age distribution for each star, and the covariance between distance modulus and log-age. We also compute the first and second moments of the effective temperature, surface gravity, metallicity, initial mass and the logarithm of the extinction. 

The pipeline will fail if $N\geq20$ meaning there is no overlap of the data with any isochrone point within $20$ times the uncertainties. In this case, the star is flagged in the output catalogue (see Section~\ref{Section::OutputCatalogue}). For each spectroscopic dataset we thin the isochrone grid in metallicity to the median metallicity uncertainty of the dataset. This results in some failures (see Section~\ref{Sec::Outputs}) which we reanalyse using the finest spacing of $0.01\dex$.

With full 6D data for the stars (on-sky position, distance, line-of-sight velocity and proper motion) we compute the Galactocentric velocity \citep[using the peculiar solar velocity from][]{Schoenrich2010}, as well as the action coordinates and guiding-centre radius \citep[using the St\"ackel fudge method,][]{Binney2012,SandersBinney2016} in the potential of \cite{McMillan2017}. 
We first draw $50$ samples from the multivariate distance modulus, line-of-sight velocity and equatorial proper motion distribution. We account for the covariance in the Gaia parallax and proper motions by first drawing samples using the Gaia measurements and then using rejection sampling to generate $50$ samples with acceptance proportional to $\mathcal{N}(\langle \mu\rangle,\sigma_\mu)/\mathcal{N}(\varpi,\sigma_{\varpi})$ where $\langle \mu\rangle$ and $\sigma_\mu$ are the mean and uncertainty in the distance modulus from our pipeline and $\varpi$ and $\sigma_{\varpi})$ are the mean and standard deviation of the parallax reported by Gaia. For each sample we compute the derived quantities and report the mean and standard deviation as the best estimate and uncertainty (ignoring unbound samples -- if no samples are bound, the resulting actions are undefined). Note that the uncertainty in the actions does not reflect the uncertainty in the solar position or the potential. 

\section{Spectroscopic datasets}\label{Sec::Datasets}
\begin{figure}
$$\includegraphics[width=\columnwidth]{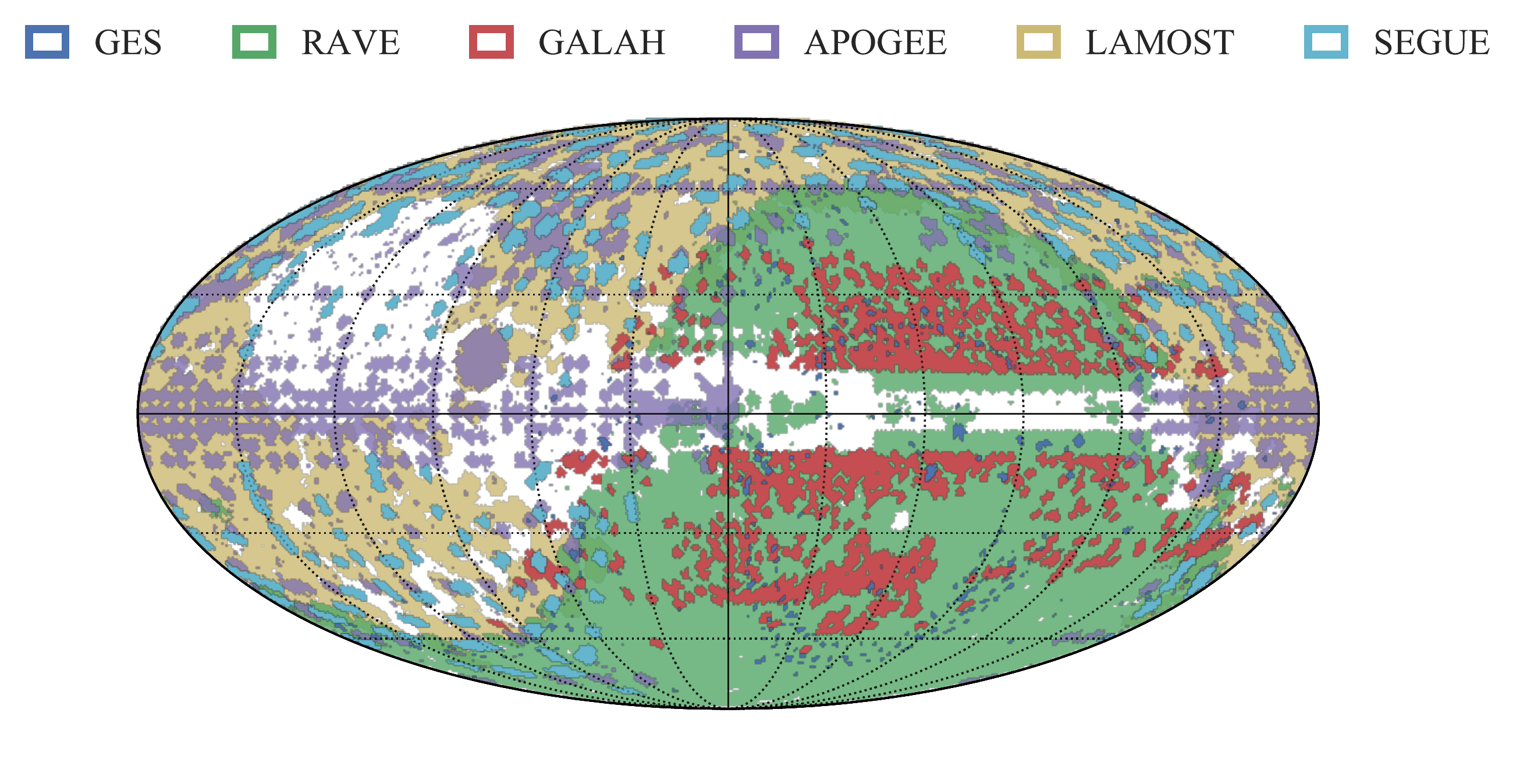}$$
\caption{On-sky galactic distribution of the stars processed by our pipeline coloured by survey.}
\label{Fig::OnSky}
\end{figure}

In this section we describe the spectroscopic datasets to which we apply our algorithm. We give a brief overview of each spectroscopic survey along with the specific data we use. In Figure~\ref{Fig::OnSky} we show the on-sky distributions of the different surveys we use. For APOGEE, LAMOST and GALAH giants, we apply the spectroscopic mass estimator. For LAMOST, we first build a data-driven model of the spectra \citep[the Cannon,][]{Casey2016} to find spectroscopic parameters calibrated to the APOGEE scale. For GALAH we build a data-driven model of the spectroscopic labels directly using a neural network to put GALAH spectroscopic parameters on the APOGEE scale.

\subsection{APOGEE}
As part of SDSS IV \citep{Blanton2017}, the Apache Point Observatory Galactic Evolution Experiment (APOGEE) has targeted primarily stars confined to the disc and bulge with infrared spectroscopy. 
The North survey uses the APOGEE 300-fibre spectrograph \citep{APOGEEspectrograph} on the Sloan 2.5m telescope \citep{Gunn2006} at Apache Point Observatory. APOGEE spectra are taken in the $H$ band ($15200\angstrom$--$16900\angstrom$) with a resolution of $R\sim22500$.
We use the APOGEE DR14 catalogue \citep{SDSSDR14} removing duplicates (retaining the highest signal-to-noise). We use the calibrated $\teff$, $\logg$ and $\mh$ \citep{GarciaPerez2016} along with the reported covariances, and the 2MASS $J,H,K_s$ photometry with uncertainties \citep{2MASS}. We use the provided radial velocities with uncertainties (stars with only a single visit have \texttt{VSCATTER}$=0$ so we assign these stars the median \texttt{VSCATTER} for all stars). Although APOGEE was designed to target giant stars, there are many nearby dwarf stars. These stars do not have provided $\logg$ in DR14 as they differed significantly from stellar models, but $\teff$ and $\feh$ is reported. For all stars with valid $\teff$ and $\feh$, but no $\logg$ we assign $\logg=4.5\pm2$. Note, we do not perform any cuts on quality of spectroscopic parameters (flagged with \texttt{ASPCAPFLAG}) instead leaving this to be done at a later stage of analysis. We use the procedure of Section~\ref{Sec::Mass} to assign masses and uncertainties given the reported $\am$, $\cm$ and $\nm$.

\subsection{LAMOST}

LAMOST \citep[Large Sky Area Multi-Object Fiber Spectroscopic Telescope,][]{Cui2012,Zhao2012} Experiment for Galactic Understanding and Exploration \citep[LEGUE,][]{Deng2012} is a low resolution ($R\sim1800$) optical ($3650-
9000 \angstrom$) spectroscopic survey designed to study the Milky Way disc and halo. From the third data release A, F, G, K catalogue (\href{http://dr3.lamost.org/}{http://dr3.lamost.org/}), we use the reported $\teff$, $\logg$ and $\feh$ (as a proxy for metallicity) and their uncertainties (approximately $3.1$ million stars). We complement the catalogue with 2MASS $J,H,K_s$ filtering on quality (different photometric catalogues were used in the LAMOST target selection). We use photometry provided:
\begin{enumerate}
\item the photometric quality flag \texttt{ph\_qual} is A, B, C or D,
\item the contamination flag \texttt{cc\_flg} is $0$.
\item if the read flag \texttt{rd\_flg} is $2$ (i.e. the magnitude is from profile fitting), \texttt{X\_psfchi} (where X is the magnitude) must be less than $3$.
\end{enumerate}
When 2MASS photometry does not satisfy these requirements, we instead use Pan-STARRS photometry ($g_P$, $r_P$ and $i_P$) along with $G$ or just $G$, $G_\mathrm{RP}$ and $G_\mathrm{BP}$ when Pan-STARRS isn't available \cite[][or potentially saturated $g_P$, $r_P$ and $i_P<13.5\,\mathrm{mag}$]{Magnier2013}. Comparisons with the Gaia RVS sample \citep{GaiaDR2RVS2} and APOGEE \cite{Anguiano2018} has demonstrated the radial velocities from LAMOST are too small by $4.5\,\mathrm{km\,s}^{-1}$ so we correct all radial velocities by this amount.

\subsubsection{Spectroscopic mass estimates}
We further complement the LAMOST giants with mass measurements using a two-stage procedure. First, we apply the Cannon method of \cite{Ho2017} to a sample of $\sim20000$ stars cross-matched between APOGEE and LAMOST (with \texttt{ASPCAPFLAG}, \texttt{C\_M\_FLAG}, \texttt{N\_M\_FLAG} $=0$). We repeat the analysis of \cite{Ho2017} as (i) we wish to use our framework of Section~\ref{Sec::Mass} which has been calibrated using updated APOGEE parameters and (ii) we are using LAMOST DR3 instead of DR2. We build a seven label ($\teff$, $\logg$, $\mh$, $\am$, $\cm$, $\nm$, $A_K$) using the DR14 calibrated APOGEE parameters using the code of \cite{Casey2017}\footnote{\href{https://github.com/andycasey/AnniesLasso}{https://github.com/andycasey/AnniesLasso}}. All spectra are normalized using a smoothed version of the spectrum (with FWHM $50\angstrom$). Unlike \cite{Ho2017}, we don't use any photometry (resulting in poorer estimates of $A_K$). 

As \cite{Ho2017} found, there is a weak correlation between the output $\am$ and line-of-sight velocity due to sky features in the spectrum. We do not correct for this by omitting regions contaminated by sky features as it does not seem a significant issue for measuring the mass (which is a weak function of $\am$). We employ a leave-$10\percent$-out scheme to measure the accuracy of our approach. The formal errors from the Cannon are significantly smaller than the scatter with respect to the test set. In a similar approach to \cite{Ho2017}, we measure the scatter $\sigma_X$ in label $X$ in bins of signal-to-noise $\mathrm{S/N}$ and fit three parameter functions of the form
\begin{equation}
\sigma_X = \frac{p_0}{(\mathrm{S/N})^{p_1}+p_2^2}.
\label{Eqn::SNrelation}
\end{equation}
We apply our model to all stars with $\logg<3.9$ and $\teff<6000\K$ in the LAMOST catalogue and deem the parameters satisfactory if the reduced chi-squared $<3$ and the results for $\teff$, $\logg$ and $\mh$ lie within the training set. We assign uncertainties using the $\mathrm{S/N}$ relation (of equation~\ref{Eqn::SNrelation}). The resulting sets of parameters ($\teff$, $\logg$, $\mh$, $\am$, $\cm$, $\nm$) are used in the mass estimator of Section~\ref{Sec::Mass} to obtain masses and uncertainties.

\subsection{RAVE}
The RAdial Velocity Experiment \citep[RAVE,][]{Steinmetz2006} was designed as a predecessor to the Gaia Radial Velocity Spectrometer (RVS). $\sim500000$ medium resolution ($R\sim7500$) spectra were taken  in the spectral range $\sim8400\angstrom$--$\sim8800\angstrom$ (around the Ca II triplet) on the 1.2 m UK
Schmidt Telescope at the Australian Astronomical Observatory
(AAO) using a multi-object
spectrograph. With a targeting magnitude range of $9<I<12$, RAVE has primarily observed local disc stars. Spectroscopic parameters were extracted from the spectra using two different pipelines. The first (denoted \texttt{RAVE\_DR5}) utilises the parameters provided in the latest RAVE data release \citep[DR5,][]{Kunder2017} using the analysis method from \cite{Kordopatis}, whilst the second (denoted \texttt{RAVE\_Cannon}) utilises the Cannon results for the RAVE spectra from \cite{Casey2017}. We remove duplicate observations retaining the higher signal-to-noise spectra. For both datasets, we use 2MASS $J,H,K_s$ photometry and associated uncertainties. For \texttt{RAVE\_Cannon} we compute the metallicity from the reported $\feh$ combined with an inverse-variance-weighted estimate of $\afe$ using the formula from \cite{Salaris1993}. We also use covariances between $\teff$, $\logg$ and $\feh$ for this dataset.

\subsection{GES}
The Gaia-ESO survey \citep[GES,][]{Gilmore2012} is a public spectroscopic survey on the Very Large Telescope (VLT) utilising the FLAMES (Fiber Large
Array Multi-Element Spectrograph) spectrograph (both medium resolution GIRAFFE $R\sim20000$ spectra and high resolution UVES $R\sim50000$). One of the goals of Gaia-ESO is to study the systematics in spectroscopic parameters by comparison of the results from multiple \emph{nodes} and through a combination of field and cluster fields. The field sample primarily focusses on thick disc and halo stars.
We use the reported $\teff$, $\logg$ and $\feh$ (as a proxy for metallicity) with the associated errors from the DR3 public data release\footnote{\href{https://www.gaia-eso.eu/data-products/public-data-releases/gaia-eso-survey-data-release-3}{https://www.gaia-eso.eu/data-products/public-data-releases/gaia-eso-survey-data-release-3}}. The target selection for field stars in GES was performed using VISTA photometry \citep{VISTA1, VISTA2} so we preferentially use VISTA photometry $J_V,H_V,K_V$ with associated uncertainties and default to 2MASS $J,H,K_s$ where unavailable (we transform the 2MASS bands provided from the PARSEC isochrones to VISTA bands using the relations from \href{http://casu.ast.cam.ac.uk/surveys-projects/vista/technical/photometric-properties}{http://casu.ast.cam.ac.uk/surveys-projects/vista/technical/photometric-properties}).

\begin{figure}
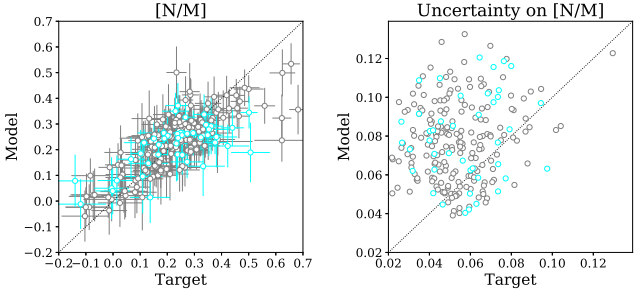

$$\includegraphics[width=\columnwidth]{{{figures/modelcomp_nitrogen}}}$$
\caption{Recovery of nitrogen abundance (left) and its uncertainty (right) from the Bayesian neural network for 193 stars in the training sample (grey) and 44 stars in the unseen testing sample (cyan) from the overlap of APOGEE DR14 and GALAH DR2 data.}
\label{Fig::NewGalahSpec}
\end{figure}

\subsection{GALAH}
GALactic Archaeology with HERMES (the High Efficiency and Resolution Multi-Element Spectrograph) \citep{DeSilva2015,Martell2017} is a medium resolution ($R\sim28000$) optical (four windows) multi-fibre spectroscopic survey. It was designed to provide a rich set of chemical abundances for the purpose of chemical tagging distinct star formation events \citep{FreemanBlandHawthorn2002} and has a simple selection function ($12<V<14$) targeting primarily disc stars. The second GALAH data release \citep{Buder2018} contains $\sim340000$ stars. The spectroscopic analysis is performed in two stages with a high-quality test-set analysed in detail with line synthesis which is then used as a training set for a Cannon model \citep{Ness2016,Casey2016}. 
As a result, there are stars that fall outside the training set producing unreliable results (flagged by \texttt{flag\_cannon}). We ignore these stars resulting in a catalogue of $264,227$. We use 2MASS magnitudes $J,H,K_s$ and compute metallicity from a combination of $\feh$ and $\afe$ using \cite{Salaris1993} (if $\afe$ is available, else we use $\feh$).

\subsubsection{Spectroscopic mass estimates}
We wish to complement the GALAH giant stars with spectroscopic mass estimates. We cannot employ the same procedure as we performed for LAMOST as the GALAH spectra are not publicly available. Instead, we build a model to put the reported spectroscopic parameters from GALAH onto the APOGEE scale. GALAH DR2 has measured $\cm$ from the spectra \citep{Buder2018}, but not $\nm$. Without $\nm$, the spectroscopic mass estimator is of limited power as the strongest correlation is between $\cn$ and mass. However, we can estimate the missing $\nm$ measurement from the other spectroscopic measurements. In particular, the conservation of total CNO in the CNO cycle produces a simple relationship between $\cm$, $\om$ and $\nm$.

We take the 237 stars that lie in the overlap between APOGEE and GALAH (\texttt{ASPCAPFLAG}$! = 2^{**}23$ and \texttt{flag\_cannon} = 0) and build a Bayesian neural network (with one hidden layer and 20 nodes) to relate the GALAH ($\teff,\logg,\mh,\cfe,\ofe$) to APOGEE ($\teff,\logg,\mh,\cm,\nm,\om$). The procedure is similar to that described in Section~\ref{Sec::Mass} except we use NUTS to find the posterior distributions of the neural network parameters on a training subsample of 193 stars, and use the remaining 44 test stars to assess the performance of the neural network.

Figure \ref{Fig::NewGalahSpec} shows how well the Bayesian neural network is able to predict the APOGEE DR14 nitrogen abundance from the GALAH DR2 spectral parameters for the training and testing samples. The model only starts failing for [N/M]$>0.5$, beyond which the model underestimates the APOGEE DR14 nitrogen abundance. Nitrogen has a negative correlation with oxygen and carbon; however the five stars with [N/M]$>0.5$ are found over the whole range of oxygen and carbon abundances. There may be a genuine physical reason underpinning this, but it's difficult to look beyond the Poisson noise obviously affecting the estimates for higher nitrogen abundances.


\subsection{SEGUE}
The Sloan Extension for Galactic Understanding and Exploration (SEGUE) \citep{Yanny2009} is a low resolution ($R\sim2000$) optical spectroscopic survey designed to complement the SDSS catalogues with radial velocities. The survey was conducted in two key stages (SEGUE-1 and SEGUE-2) where SEGUE-2 focussed primarily on the outer halo of the Galaxy. Spectroscopic parameters have been computed by the SEGUE Stellar Parameter Pipeline (SSPP) \citep{Lee2008a,Lee2008b,AllendePrieto2008,Smolinski2011}. We adopt the external error estimates from \cite{Lee2008a} of $141\K$, $0.23\dex$ and $0.23\dex$ in $\teff$, $\logg$ and $\feh$ respectively which we add in quadrature to the reported uncertainties. We take all SDSS DR12 stars observed in the programmes SEGUE, SEGUE-2 or SEGUE-faint with valid $\teff$, $\logg$ and $\feh$, that are science primary, have \texttt{zwarning} $ = 0$ or $16$ and are flagged as normal (`nnnnn'), resulting in a catalogue of $187,152$ stars). We use $\feh$ as a proxy for metallicity. We complement with the $g,r,i$ photometry from SDSS.

\subsection{Complementary Gaia data}

For each catalogue, we perform a $5\arcsec$ radius cross-match to the second Gaia data release \citep{GaiaDR2} 
by utilising the Gaia proper motions and accounting for the respective epochs of the surveys (we assume the epoch of all the spectroscopic catalogue observations is 2000). 
From the Gaia DR2 source catalogue, we extract the parallax, proper motion and the uncertainty covariance matrix for the astrometry. Despite the reported global zero-point parallax offset of $\sim30\,\mu\mathrm{as}$ \citep{GaiaDR2Astrometry,Helmi2018}, we use the reported parallax and uncertainty as is. For studying large-scale Galactic structure, the zero-point is only important for more distant stars where other uncertainties (e.g. in the choice of prior) are also significant. Additionally, the zero-point is also a strong function of on-sky location, magnitude and colour \citep{GaiaDR2Astrometry,Helmi2018,Riess2018} so a global offset will only fix some systematic issues. For the Gaia $G$ photometry we assume a $0.02\,\mathrm{mag}$ systematic uncertainty floor in the photometry that reflects systematics in the photometry \citep{GaiaDR2Photometry1} as well as intrinsic uncertainty in the isochrones. We use the parallax and $G$ photometry as additional inputs in our pipeline. When the cross-match fails, we still process the stars without any Gaia data.

\subsection{Output catalogue description}\label{Section::OutputCatalogue}

\begin{figure}
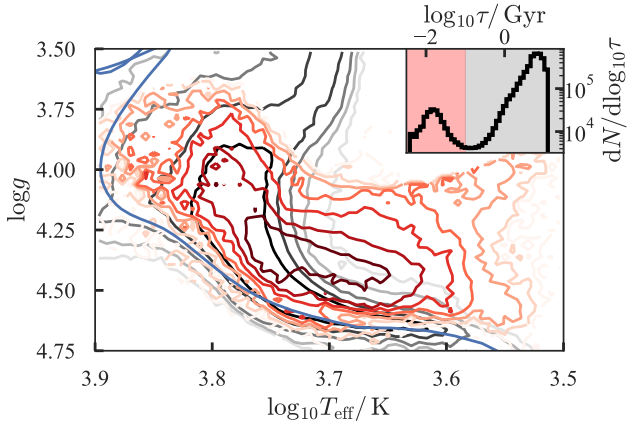

$$\includegraphics[width=\columnwidth]{{{figures/age_binary}}}$$
\caption{Spectroscopic HR diagram of stars assigned a young age by our pipeline: black contours show the distribution of stars with $\tau>100\Myr$ and red $\tau<100\Myr$ which are suspected binaries. The inset shows the distribution of $\log_{10}\tau$ shaded by the separation. The blue line shows a $1\Gyr$ solar metallicity isochrone.}
\label{Fig::AgeBinary}
\end{figure}

We provide a catalogue of the combined results from all surveys complete with the Gaia DR2 \texttt{source\_id} for each entry (where available)\footnote{The full catalogue is available at \href{https://www.ast.cam.ac.uk/~jls/data/gaia_spectro.hdf5}{https://www.ast.cam.ac.uk/\~jls/data/gaia\_spectro.hdf5}}. We provide a full description of the columns (including their units) in Table~\ref{Table::Catalogue}. For each survey, we include a unique identifier (APOGEE: \texttt{APOGEE\_ID}, LAMOST: \texttt{obsid}, RAVE: \texttt{raveid}, GES: \texttt{CNAME}, GALAH: \texttt{sobject\_id}, SEGUE: \texttt{specobjid}) and we provide a field \texttt{survey} with a string detailing which survey the entry comes from. Note that we have not removed duplicate stars that were observed by two separate surveys (e.g. RAVE and RAVE-On). For each star we include the cross-matched Gaia DR2 \texttt{source\_id}, the angular separation of the cross-match and the photometric bands used in the pipeline. The table is saved as an \texttt{astropy} table \citep{astropy} in \texttt{hdf} format, and can be simply read with \texttt{from astropy.table import Table; data = Table.read('file');}. A description of the column and structure of the table is available in \texttt{data.meta['COMMENT']}.

We provide a \texttt{flag} entry in the table. If the isochrone pipeline has failed (e.g. there is no overlap between isochrones and data within $20$ times the uncertainties due to a bad cross-match), \texttt{flag}=1. This flag is also non-zero if there is a problem with the input spectroscopy (\texttt{flag}=2), photometry (3), astrometry (4) or mass (5) (no entries in the catalogue have \texttt{flag}=4,5).
A small fraction ($\sim2\percent$) of the processed stars appeared to have overlap with only a single isochrone (point) leading to zero or undefined 2nd moments (uncertainties). We flag these stars with $\texttt{flag}=6$. 

We found a non-negligible fraction of processed red stars had small ages. In Fig.~\ref{Fig::AgeBinary}, we show the 1D distribution of $\log_{10}\tau$ which exhibits a clear peak at $-2$. We also display the $\teff$-$\logg$ distribution of these stars which reveals that the pipeline has assigned these as pre-main-sequence objects. The assigned $\teff$ and $\logg$ for these stars form a clear sequence running alongside the main sequence, such that it is highly likely that many of these stars are binary. This is also evident in colour-magnitude space. We flag the stars for which $\tau<100\Myr$ and $\logg>3.5$ and $\log_{10}\teff<3.9$ with $\texttt{flag}=7$. For more accurate parameters, these stars require a special pipeline such as that presented in \cite{Coronado2018}.

Some surveys contain duplicate observations of the same stars and some surveys have observed the same stars more than once. We identify duplicates using the Gaia source IDs and preferentially keep the entries flagged as $\texttt{flag}=0$ from different surveys in the order: APOGEE, GALAH, GES, RAVE-ON, RAVE, LAMOST and SEGUE, and for duplicates within surveys we keep the result with the smallest vertical velocity uncertainty (to retain both the most precise distance estimates [from the spectroscopy] and the most precise radial velocities).
We provide a \texttt{duplicated} flag which is $1$ for the rejected duplicates and zero otherwise.  Finally, we provide a \texttt{best} flag which is $1$ if \texttt{flag}=0, \texttt{duplicated}=0 and the star has a Gaia match. 
We provide statistics for all processed stars in Table~\ref{Table::Numbers}. Our `best' sample consists of $\sim3$ million stars. It should be noted that additional quality cuts may be necessary for certain analyses to remove stars that have been flagged as unusual (for instance, in RAVE stars with \texttt{c$^*$}!=`n').

\begin{table*}
\caption{Numbers of stars processed by pipeline. \emph{Total} gives the total number of stars in the survey, \emph{Gaia matches} gives the number of matches in Gaia DR2, and \emph{Success} gives the number of stars with \texttt{flag}=0. The middle section shows the number of stars with pipeline failure (along with the allocated flag in brackets). \emph{Success with Gaia} gives the number of stars with \texttt{flag}=0 with Gaia DR2 matches. In the \emph{All no dupl.} column we only consider Gaia matches and have removed duplicate Gaia entries.}
\begin{tabular}{llllllllll}
\hline
 Survey                & RAVEDR5   & RAVEON & GALAH  & APOGEE & LAMOST  & GES   & SEGUE  & All     & All no dupl. \\
 Total                 & 457555 & 457555 & 342682 & 258475 & 3177995 & 25332 & 187152 & 4906746 & 3706733      \\
 Gaia matches          & 456353 & 456353 & 342212 & 256851 & 3168545 & 25313 & 187100 & 4892727 & 3706733      \\
 Success (0)           & 415200 & 376316 & 260233 & 203417 & 2861310 & 10882 & 182132 & 4309490 & 3318119      \\
  \hline
 Pipeline failed (1)   & 1358   & 368    & 67     & 2778   & 15503   & 105   & 393    & 20572   & 16928        \\
 Spec. problem (2)     & 20602  & 67439  & 78455  & 42553  & 67524   & 12513 & 0      & 289086  & 185066       \\
 Phot. problem (3)     & 498    & 498    & 0      & 810    & 7660    & 1181  & 1490   & 12137   & 4667         \\
 Unreliable errors (6) & 3220   & 1093   & 46     & 460    & 7948    & 52    & 72     & 12891   & 6444         \\
 Low age (7)           & 16677  & 11841  & 3881   & 8457   & 207577  & 599   & 3065   & 252097  & 166986       \\
  \hline
 Success with Gaia     & 414238 & 375488 & 259877 & 203127 & 2858287 & 10881 & 182087 & 4303985 & 3318119      \\
\hline
\end{tabular}
\label{Table::Numbers}
\end{table*}

\textbf{Caveats:} Our provided catalogue has a number of caveats and features that we should highlight. First, we note that when inspecting and analysing this catalogue one should be aware of our choice of Galaxy prior. This choice imposes some level of structure on the results. When attempting to fit the models to the provided dataset, this must be accounted for by `dividing out' our choice of prior.
We are using carbon and nitrogen abundances to inform giant age estimates which requires good knowledge of the initial carbon and nitrogen abundance of each star before any processing and dredge-up. Although we have tried to only apply this procedure to stars within the convex hull of our training set, it may be inappropriate for more extreme populations e.g. accreted halo stars.
When inspecting the catalogue in age and metallicity, the imprint of the isochrone gridding is visible with a weak preference fo stars to bunch at the isochrone points. Some of these stars will have very small uncertainty in age and metallicity and so in practice we recommend a minimum age uncertainty be used. The minimum spacing in metallicity and log-age of the isochrones is $0.01\dex$ and $0.05\dex$ respectively, although from inspecting the age uncertainties discreteness effects are only visible when $\sigma_{\log_{10}\tau}\lesssim0.015\dex$. Finally, our choice of prior restricts distant and metal-poor stars to be old and no star can be older than our final isochrone point. This biases the results for large ages and there is a correlation between uncertainty and age for these old stars.

Our catalogue could be further improved with additional data and modelling improvements. For instance, we have not explicitly included $\am$ abundance which is now regularly provided by large spectroscopic surveys. Furthermore, with access to the stellar spectra (from for instance RAVE and GALAH), we could improve age estimates from the giant stars (using $\nm$ or CN bands in the case of RAVE). Finally, the hard truncation in age and our adopted prior produces a number of undesirable features which could be improved through further refinements.

\section{Results}\label{Sec::Results}
Our catalogue represents the largest, most homogeneous catalogue of distances, ages, masses and spectroscopic parameters available. In this section, we demonstrate the properties of the catalogue, give a number of checks of its quality and highlight its possible power in studies of the dynamical structure of the Galaxy.

Our catalogue has three key advantages: 1. we constrain the uncertain distances from Gaia parallaxes alone for distant stars using spectroscopy and photometry, 2. the extinction estimates use spectroscopy to better pin down the intrinsic spectral type and 3. we provide age estimates, crucial for Galactic archaeology. There are two key subsamples of stars that have precise ages from our pipeline. First, the combination of parallaxes and spectroscopic metallicities break the metallicity-age degeneracy for turn-off stars \citep[e.g.][]{Howes2018}. Secondly, for giant stars accurate ages are possible due to the employed spectroscopic mass estimates combined with Gaia parallaxes which further constrain the luminosity and hence age. We conservatively define these two subsamples as 1. giants: $\logg<3\dex$ and $\log_{10}(T_\mathrm{eff}/\,\mathrm{K})<3.73$, and 2. turn-off: $3.6<\logg<4.5\dex$ and $\log_{10}(T_\mathrm{eff}/\,\mathrm{K})<4.1$.

\begin{figure}
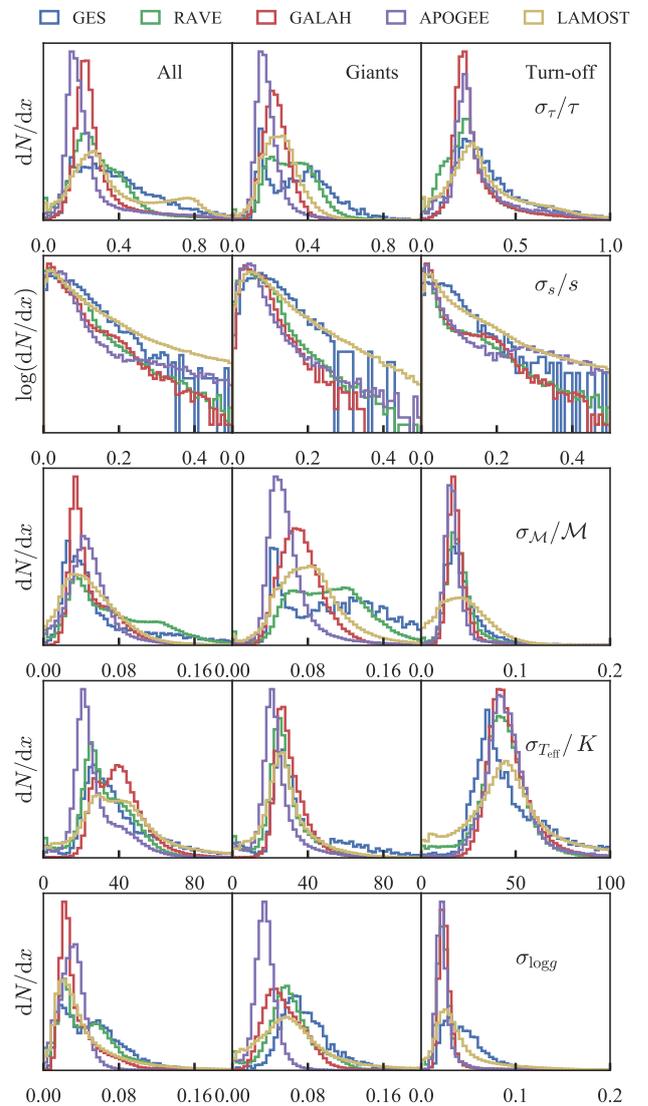

$$\includegraphics[width=\columnwidth]{{{figures/uncertainties}}}$$
\caption{Distribution of output uncertainties from our pipeline: each row corresponds to a different quantity labelled in the right plots, and each column to a different sample (defined in the text) labelled in the top panels.}
\label{Fig::uncertainties}
\end{figure}

\subsection{Output uncertainties}

In Figure~\ref{Fig::uncertainties}, we show the output uncertainties from our pipeline for all stars, the giant stars and the turn-off stars. We opt not to show the results for SEGUE as the majority of these stars are distant and metal-poor so the uncertainties (particularly in age) strongly reflect the prior. We see that for giant stars we obtain uncertainties in age of $\sim16\percent$ 
for APOGEE (which has the most accurate spectroscopic mass estimates), $\sim21\percent$ for GALAH (for which we have inferred $\nm$ from other abundances before computing a spectroscopic mass estimate) $\sim25\percent$ for LAMOST (which also has spectroscopic mass estimates) and $\sim40\percent$ for RAVE and GES (for which no spectroscopic mass estimates are used). However, there are a number of metal-poor distant giants in these surveys that have $\sim20\percent$ mass estimates due partly to the age prior employed. The mass uncertainties for the giant stars are $\sim 1/3$ the age uncertainties. For the turn-off stars, all surveys yield ages accurate to $\sim25\percent$ with GES and LAMOST producing a slightly fatter tail to large uncertainties due to larger metallicity errors.

The distribution of distance errors peaks very close to zero due to the accuracy of the Gaia parallaxes. For the giant stars (which are at larger distance) the peak moves to higher $\sigma_s/s$. The tails of the relative distance uncertainty distributions are a combination of the different survey selection functions and the quality of the input parameters. We see the largest errors arise from LAMOST and GES (which observe the faintest stars).

Our pipeline takes as inputs the spectroscopic parameters, $\teff$ and $\logg$, but also provides these as outputs. In the lower panels of Figure~\ref{Fig::uncertainties}, we show the output uncertainties in these parameters. For all surveys, we find uncertainties in $\teff$ of $20-30\,\mathrm{K}$ and $\sim50\K$ for giants and turn-off stars respectively. The effective temperature is strongly constrained with photometry, hence the independence with survey. However, the degree to which the output uncertainties in effective temperature can be trusted is unclear as they are dependent on systematics in the bandpasses and the assumed extinction law. For $\logg$, the uncertainties are survey dependent -- we see the two surveys with spectroscopic mass estimates for giants (APOGEE and LAMOST) peak at smaller $\logg$ uncertainties ($\sim0.04,0.06\dex$ respectively) than the other surveys ($\sim0.08\dex$). For the turn-off stars, the difference in $\logg$ accuracy is due to the selection functions of the surveys, where LAMOST and GES target more distant stars with less accurate distances from Gaia and hence less accurate $\logg$.

\begin{figure}
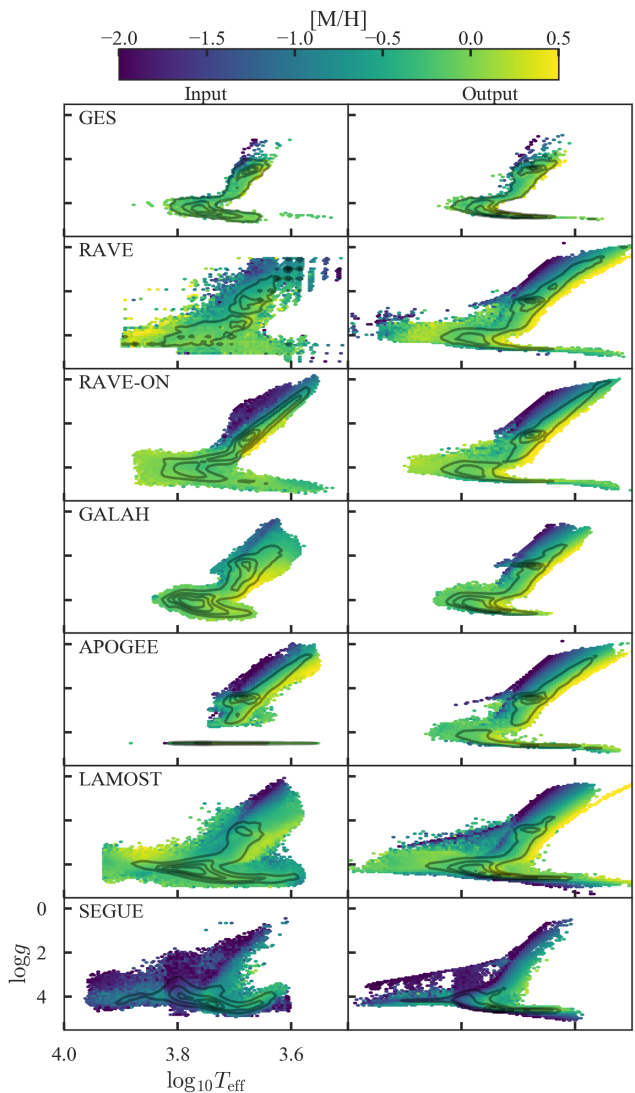

$$\includegraphics[width=\columnwidth]{{{figures/hr_diagrams_hex}}}$$
\caption{Input and output spectroscopic parameters ($\teff$ against $\logg$ coloured be $\mh$). The left column shows the spectroscopic parameters used as inputs in the pipeline and the right column the parameters output by our pipeline. Each row corresponds to a different survey labelled in the left plot. The colours correspond to the median metallicity (provided there are at least two stars in a bin) and the contours are spaced equally in the square-root of the number of stars in each bin.}
\label{Fig::HRdiagrams}
\end{figure}

The output $\teff$ and $\logg$ from our method can be compared with the inputs determined entirely spectroscopically. We show a comparison of the approaches in Figure~\ref{Fig::HRdiagrams}. We clearly see the power of enforcing an isochrone prior on the spectroscopic parameters. In particular, the main sequences are very tight due to the quality of the Gaia parallaxes. We also note that our procedure has introduced some discreteness to the diagrams due to our chosen isochrone spacing. However, each parameter estimate also has an associated error that smooths over this discreteness.

\subsection{Extinction maps}
\begin{figure}
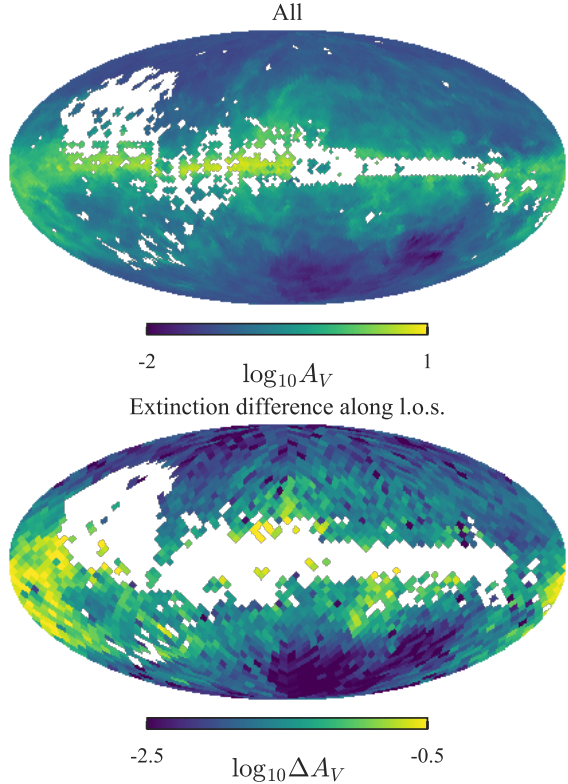

$$\includegraphics[width=\columnwidth]{{{figures/extinction_vertical}}}$$
\caption{Extinction maps in galactic coordinates for $V$ band extinction, $A_V$. The top panel shows all stars with \texttt{best}=1, and the bottom panel shows the logarithm of the difference in median $A_V$ in each bin between a slice with $0.15<s/\kpc<0.25$ and a slice with $0.25<s/\kpc<0.35$.}
\label{Fig::ExtinctionMaps}
\end{figure}

The combination of photometric, spectroscopic and astrometric data is powerful for mapping the 3D dust extinction throughout the Galaxy. Spectroscopy allows us to identify intrinsically similar stars, photometry gives a measure of the differential reddening between these stars, and astrometry tells us how the reddening varies along the line-of-sight. $A_V$ measurements for all stars are an output of our pipeline. Our measurements are informed by the adopted extinction prior but stellar colours and effective temperatures provide additional information producing stronger constraints on the extinction than provided by the prior. In Fig.~\ref{Fig::ExtinctionMaps}, we show on-sky $A_V$ for all stars and the logarithm of the difference in median extinction between two distance slices of $0.15<s/\,\mathrm{kpc}<0.25$ and $0.25<s/\,\mathrm{kpc}<0.35$ for the `best' dataset. We see the expected large-scale dust structure and the extinction increasing as a function of distance. The combination of both distances and extinctions allow full 3D extinction maps of the Galaxy to be built.

We also compare our $A_V$ measurements with $A_G$ provided in the Gaia DR2 source catalogue, calculated using a combination of the Gaia photometry and parallaxes. We see that above $A_G\approx0.3$ there is a clear linear relationship between our output $A_V$ and $A_G$ with approximate gradient $0.77$ in agreement with the expectation from Table~\ref{Table::Monochromatic}. Below $A_V\approx0.3$, $A_G$ is poorly constrained, possibly due to limitations in the methodology \citep{Andrae2018}.


\begin{figure}
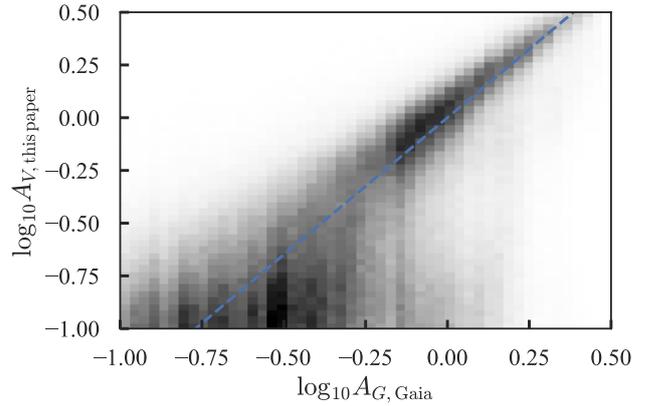

$$\includegraphics[width=\columnwidth]{{{figures/extinction_ag_av}}}$$
\caption{
Comparison of extinction estimates $A_V$ to Gaia DR2 catalogue results (2d histogram is log-scaled). The line has gradient $0.77$ equal to the anticipated value for red stars from Table~\ref{Table::Monochromatic}.
 }
 \label{Fig::ExtinctionComparison}
\end{figure}

\subsection{Age maps}
\begin{figure*}
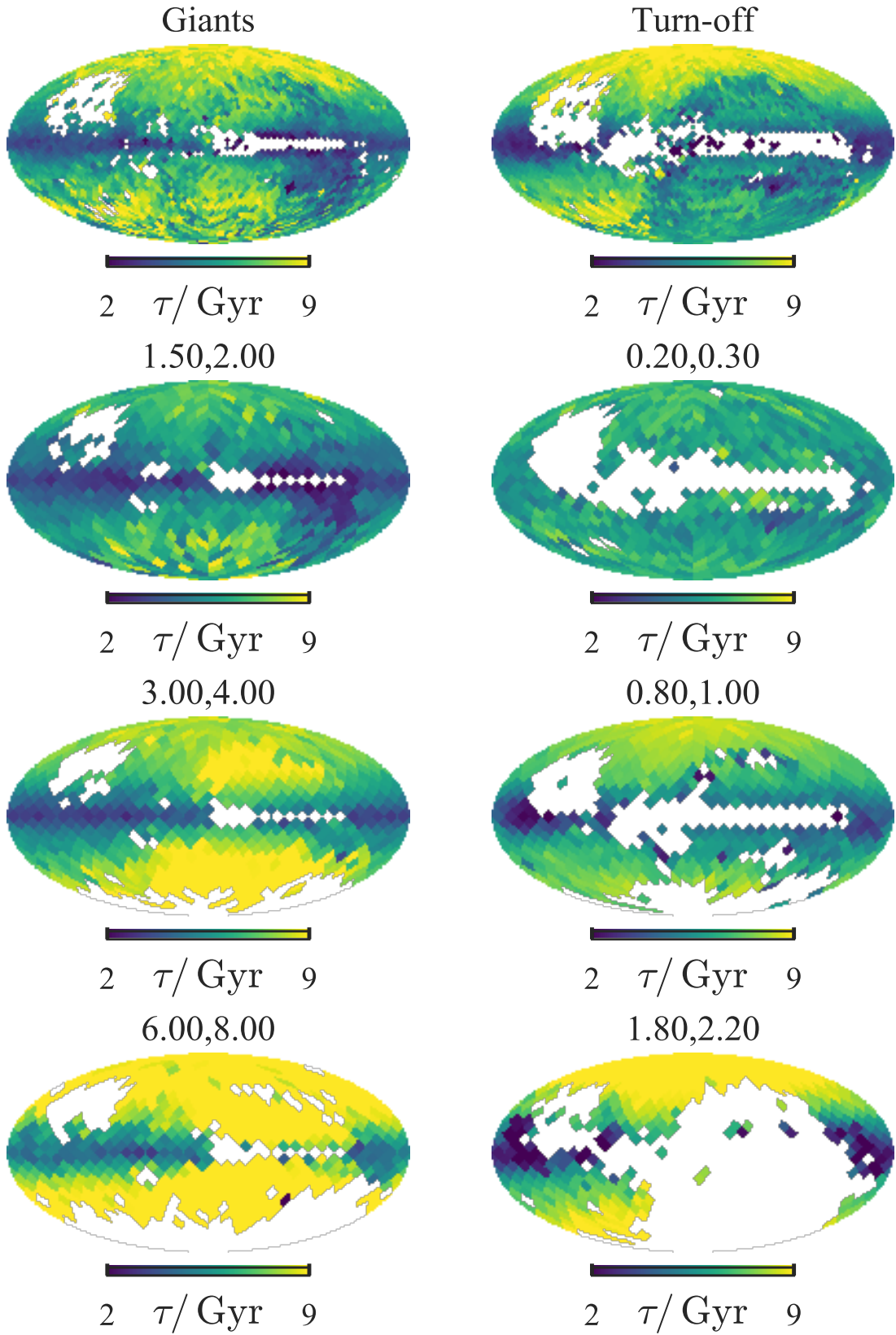

$$\includegraphics[width=0.9\textwidth]{{{figures/age_distribution_split}}}$$
\caption{All-sky age maps in galactic coordinates: left panels show giant stars and right panels turn-off stars. Each row corresponds to different distance brackets (shown in $\kpc$) above the plots with the top panels showing all stars. The different survey selection functions are clearly imprinted on the top right image.}
\label{Fig::ages}
\end{figure*}

In Figure~\ref{Fig::ages} we present all-sky age maps for the giant and turn-off subsamples. The entire giant subsample exhibits the expected gradients with galactic coordinates with younger stars confined primarily to the disc plane and higher latitudes dominated by older stars. As we slice through in distance we see the increasing contribution of old stars at high latitude (due in part to our age prior). 

The turn-off subsample exhibits a less clean picture of on-sky age distribution and for the entire sample some of the observed features are produced by the different selection functions of the surveys. In particular, the southern sky surveys (GALAH and RAVE) observe brighter stars than the primary northern sky survey (LAMOST). If we restrict ourselves to observing the nearby bright turn-off stars ($0.2<s/\kpc<0.3$) we find that there is essentially no on-sky age gradients. Over these distance scales, all age populations are contributing so we end up observing the average age of the solar neighbourhood. For a more distant bin ($0.8<s/\kpc<1$) we observe the anticipated age gradient with latitude and the southern sky surveys smoothly match into the northern sky. For an even more distant bin ($1.8<s/\kpc<2$), we lose southern sky coverage in the turn-off sample and are dominated by LAMOST which shows very strong age gradients on these large Galactic scales.

\subsection{Age-kinematic relations}
As a demonstration of the power of our catalogue, we briefly investigate some age-kinematic correlations in the disc. Studies of disc populations subdivided by age (or $\am$) \citep[e.g.][]{Bovy2016,Mackereth2017} have tended to be restricted to spatial-chemical correlations due to the limitations of pre-Gaia proper motions. Our catalogue opens up the possibility of inspecting not just spatial-chemical correlations but full chemo-kinematic correlations. We reserve a full analysis to a future project and instead here give a flavour of what is possible.

We compute the velocity dispersion of populations separated into radial and age bins. When computing the velocity dispersion in each bin, we account for the uncertainty in each velocity measurement $\sigma_{vi}$ in the following way. We first perform a $3\sigma$ clip of the raw velocities $v_i$ to remove outliers and halo contaminants. We then seek to maximise the log-likelihood
\begin{equation}
\ln\mathcal{L}=-\frac{1}{2}\sum_i\Big(\frac{(v_i-m)^2}{\sigma^2+\sigma_{vi}^2}+\ln(\sigma^2+\sigma_{vi}^2)\Big),
\end{equation}
which can be found by solving
\begin{equation}
0=\sum_i\Big(\frac{(v_i-m)^2}{(\sigma^2+\sigma_{vi}^2)^2}-\frac{1}{\sigma^2+\sigma_{vi}^2}\Big).
\end{equation}
Here $m$ is the simple mean computed without accounting for the uncertainties. We find the value of $\sigma$ that solves this equation by Brent's method. The uncertainty in the resulting estimate can be found from the second derivative of the log-likelihood and is approximately $\sigma\sqrt{1/2N}$ for $N$ stars. We consider at least $50$ stars per bin resulting in dispersion uncertainties better than $10\percent$.

\begin{figure}
$$\includegraphics[width=\columnwidth]{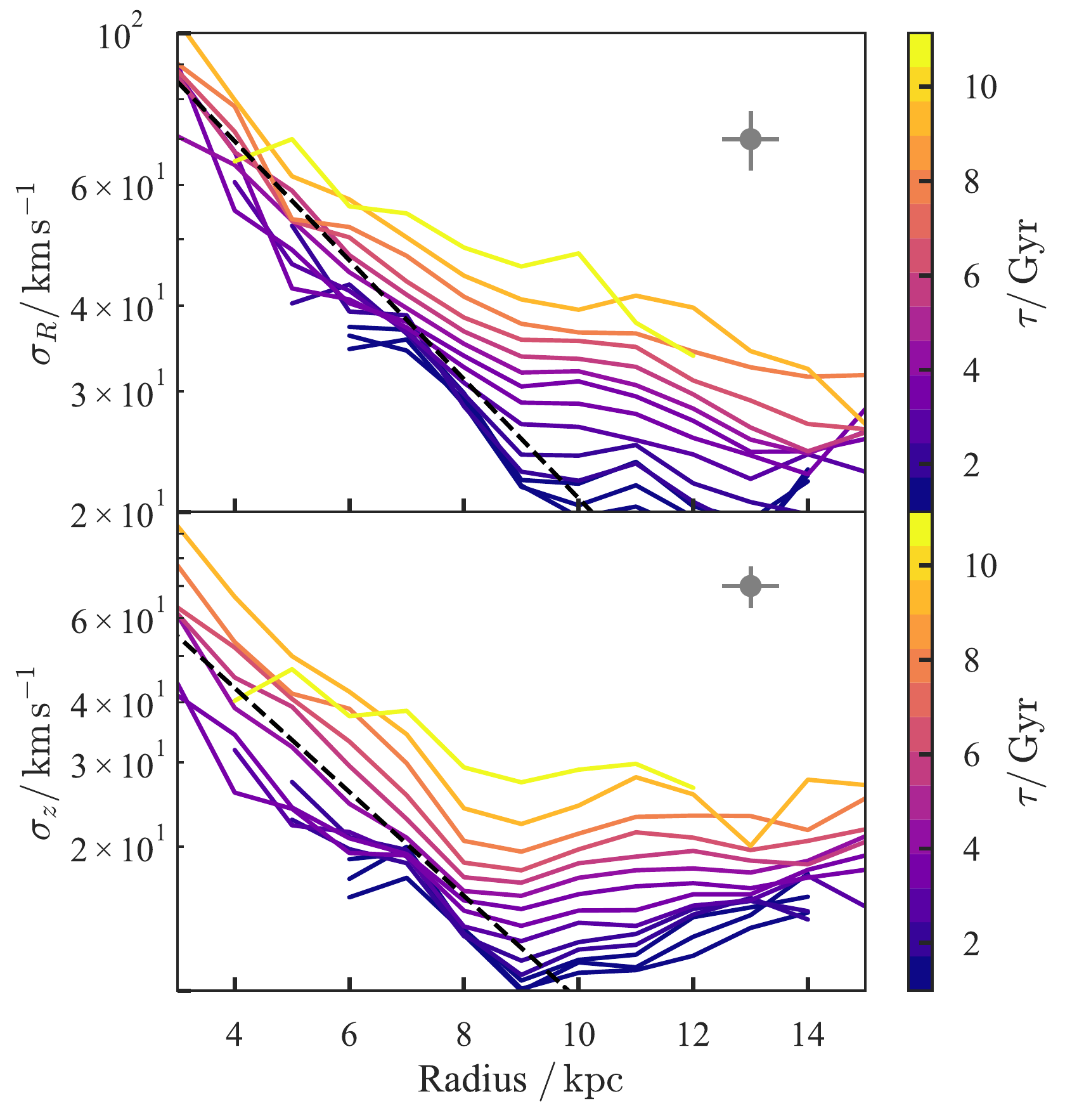}$$
\caption{Velocity dispersions against Galactocentric radius for a series of age bins (for giant and turn-off stars with $|z|<0.6\kpc$ and $\mh>-1\dex$). Each line is coloured by the mean age of the bin. The top panel shows the radial velocity dispersion and bottom panel the vertical dispersion. The dashed black lines are to guide the eye and correspond to exponentials with scale radii of $5\kpc$ and $4\kpc$ respectively. The $y$ errorbar show the maximum error for each datapoint (the $x$ errorbar is the bin width).}
\label{Figure::SigmaRadius}
\end{figure}
\begin{figure}
$$\includegraphics[width=\columnwidth]{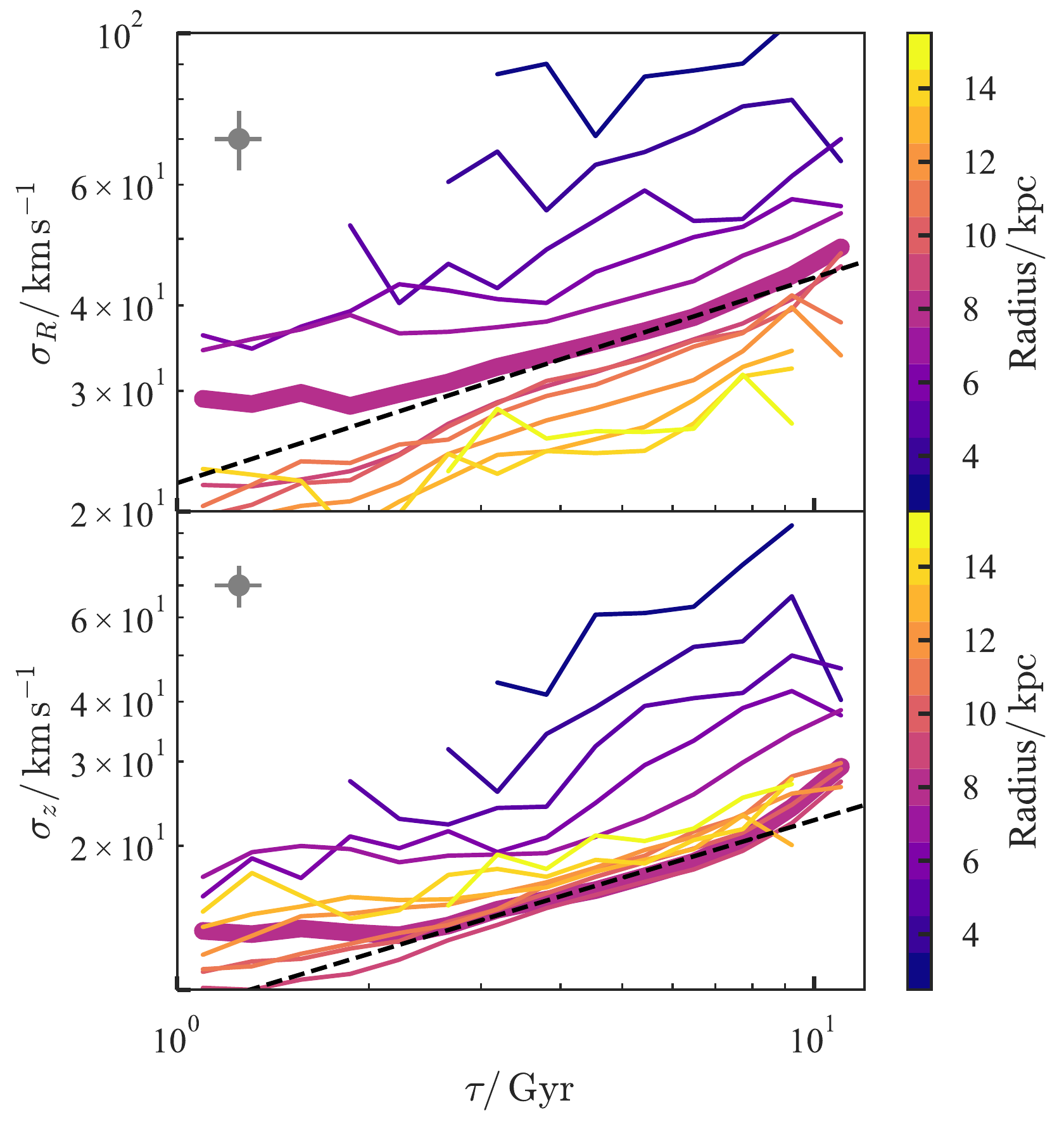}$$
\caption{Velocity dispersions against age for a series of radial bins (for giant and turn-off stars with $|z|<0.6\kpc$ and $\mh>-1\dex$). Each line is coloured by the mean Galactocentric radius of the bin. The top panel shows the radial velocity dispersion and bottom panel the vertical dispersion. The thick line corresponds to bin centred around the solar neighbourhood. The dashed black lines are to guide the eye and correspond to power laws $\tau^\beta$ with coefficients $\beta=0.3$ and $\beta=0.4$ respectively. The $y$ errorbar show the maximum error for each datapoint (the $x$ errorbar is the bin width).}
\label{Figure::SigmaAge}
\end{figure}

In Figure~\ref{Figure::SigmaRadius} we give the Galactocentric radial dispersion profile of the radial and vertical velocity separated into different age bins. We use all `best' giant and turn-off stars with $|z|<0.6\kpc$, $\mh>-1\dex$ and $\sigma_\tau/\tau<0.45$ (approximately $1.2$ million stars). Inwards of the solar radius, both velocity dispersions decline approximately exponentially with scale radii of $\sim5\kpc$. Near and beyond the solar radius, the velocity dispersion profiles flatten with radius. The radial velocity dispersion continues to decline whilst the vertical dispersion plateaus and begins to weakly increase. This intriguing signature could be linked to the theoretical claims that mono-age populations show significant flaring \citep{Minchev2015}, although it is also possibly due to the selection effect of seeing higher latitude stars at larger Galactocentric radii. A similar flattening of the vertical velocity dispersion profile is seen for the Gaia RVS sample studied by \cite{GaiaMWKinematics}. At all radii, we clearly see a gradient in age with lower age populations on kinematically colder orbits. The break in the behaviour of the velocity dispersion profiles occurs at decreasing radius with increasing age. A flattening of velocity dispersion can be caused by distance systematics which for the sample we consider more strongly affect the vertical dispersion. 
A reasonable parallax systematic of $\sim0.03\,\mathrm{mas}$ \citep{GaiaDR2Astrometry} $\sim3\,\mathrm{km\,s}^{-1}$ $7\kpc$ away and so is unlikely to fully explain the effect seen in Fig.~\ref{Figure::SigmaRadius}. Additionally, the spectroscopy and photometry used in our distance measurements goes some way to correcting any Gaia parallax systematic issues at larger distances.

The behaviour of the velocity dispersions with age can be observed more clearly in Figure~\ref{Figure::SigmaAge} where we show the dispersion profiles with age split by Galactocentric radius. At all radii, the velocity dispersions grow approximately like a power-law with age, $\tau^\beta$. We find coefficients of $\beta\sim0.3$ for the radial dispersion and $\beta\sim0.4$ for the vertical dispersion although this is not accounting for the (significant) age uncertainties which act to flatten the relations \citep{Martig2014, AumerBinney2016b}. However, these values are consistent with observations of the solar neighbourhood \citep{AumerBinney} and indicate the velocity dispersions throughout the Galaxy are consistent with heating by a combination of spiral arms and molecular clouds \citep{AumerBinney2016a}. With our particular set of cuts (in particular, only considering low Galactic height stars), we find no real indication of a break in the dispersion profile at intermediate/large age. Inside the solar radius, there is the suggestion of the profiles flattening with age (although this is possibly due to increased age uncertainties). We note that our observations must all be considered within the context of the selection function of the sample. A future work will account for these effects in a full model of our presented catalogue, enabling us to disentangle observational biases from features of the Galaxy.

\begin{figure*}
$$\includegraphics[width=\textwidth]{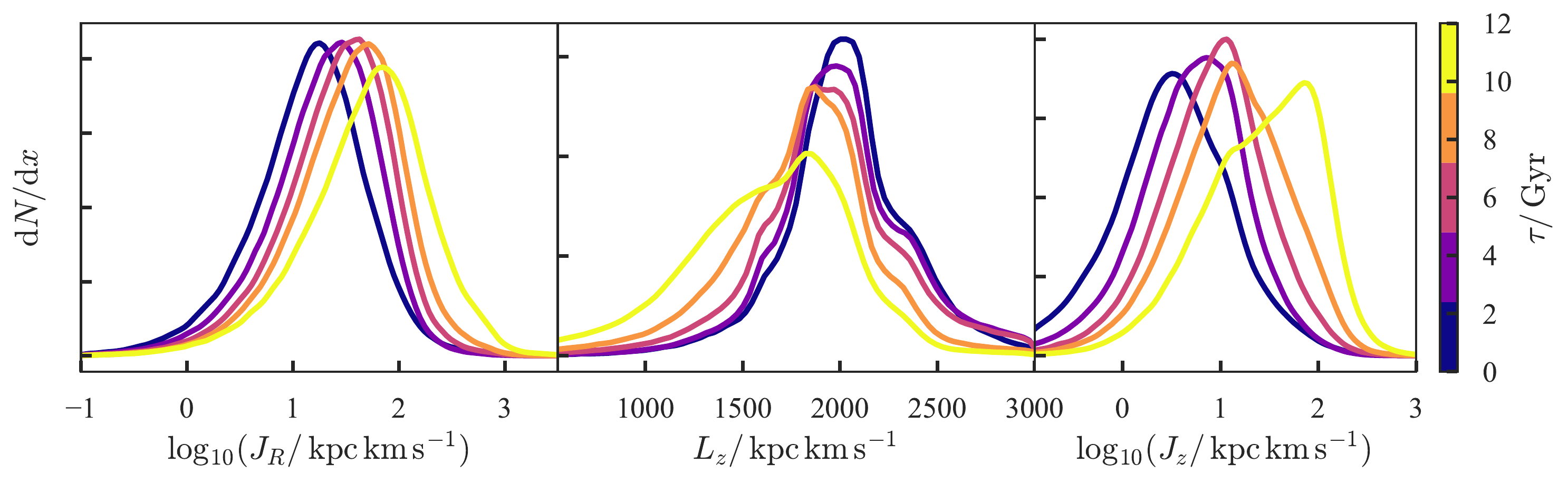}$$
\caption{Normalized kernel density estimates for the action distributions in six evenly-spaced age bins from $1$ to $12\Gyr$ (for giant and turn-off stars with $|z|<2.5\kpc$ and $\mh>-1\dex$) -- left panel shows logarithm of the radial action, central panel the $z$-component of the angular momentum and right panel the logarithm of the vertical action.}
\label{Figure::ActionDist}
\end{figure*}

In our output catalogue, we also provide estimates of the actions (with associated errors). In Figure~\ref{Figure::ActionDist} we show the action distributions (estimated using a kernel density estimate) split into six age bins using all `best' giant and turn-off stars with $|z|<2.5\kpc$, $\mh>-1\dex$ and $\sigma_\tau/\tau<0.45$. 
Mirroring the results from Figure~\ref{Figure::SigmaRadius}, we see the steadily increasing mean radial and vertical action with increasing age. We observe that the oldest age bin in vertical action is signifcantly skewed indicating the presence of a thick disc component. This is not mirrored in the radial action distribution. The picture from the radial action distributions alone is of a smooth quiescent evolution of the disc.
The $z$-component of angular momentum smoothly declines with age reflecting both the asymmetric drift in the populations and the spatial distribution due to inside-out growth. The angular momentum distribution also has more evidence of the selection function of the catalogue (as $z$-component of angular momentum is a proxy for radius for dynamically cool stars).



\section{Conclusions}\label{Sec::Conclusions}

We have presented a catalogue\footnote{The catalogue is available from \href{https://www.ast.cam.ac.uk/~jls/data/gaia_spectro.hdf5}{https://www.ast.cam.ac.uk/~jls/data/gaia\_spectro.hdf5} and the corresponding code at \href{https://github.com/jls713/gaia_dr2_spectro}{https://github.com/jls713/gaia\_dr2\_spectro}. The format of the catalogue is given in Table~\ref{Table::Catalogue}.} of approximately $3$ million ages, masses, distances, extinctions and spectroscopic parameters for stars in common with large spectroscopic surveys and the second Gaia data release. We considered the results from APOGEE, LAMOST, SEGUE, Gaia-ESO, RAVE (using both the RAVE DR5 results and RAVE-On) and GALAH giving approximate all-sky coverage. We have complemented our catalogue with estimates of Galactocentric coordinates and actions along with associated (one-dimensional) uncertainties. As well as presenting details of our procedure, we have focussed on the quality of the output catalogue and presented some preliminary results demonstrating its power. Our conclusions are as follows.
\begin{enumerate}
\item A non-negligible fraction of our catalogue is assigned pre-main sequence properties from our pipeline. Many of these stars lie along the binary sequence in colour-magnitude space so are suspected binary stars. Our catalogue can be used to assign binarity to the stars observed by the considered spectroscopic surveys.
\item We have investigated the output uncertainties produced by our catalogue. Two key subsamples for studying the age structure of the Galaxy are giant stars and turn-off stars. For the APOGEE, LAMOST and GALAH subsamples, we have employed techniques to assign spectroscopic mass estimates to the giant stars (via carbon and nitrogen abundances). This results in ages accurate to $15-25\percent$. For turn-off stars, parallax is insufficient to determine an accurate age due to a degeneracy with metallicity. Employing spectroscopic metallicity measurements results in output uncertainties of $\sim20-30\percent$. We also have provided output spectroscopic parameters $\logg$ and $\teff$ which typically have uncertainties less than $0.1\dex$ and $80\K$ respectively. These parameters can be used as initial guesses in improved analysis of the spectra.
\item We have provided output extinction estimates for all stars in our catalogue. These estimates are informed by prior extinction maps but the output uncertainties are typically smaller than the prior input uncertainties. When complemented with our output distances, the extinction values can be used to construct a 3D extinction map (over the volume probed by our spectroscopic samples).
\item We presented some first results on the correlations between kinematics and age for Milky Way stars using our new catalogue. We demonstrated that the Galactocentric radial profile of radial and vertical velocity dispersions appears to flatten beyond the solar radius. At all radii, we see a smooth power law increase of the (radial and vertical) velocity dispersions with age.
\end{enumerate}

The presentation of our catalogue represents a first step in analysing the chemo-dynamic composition of the Galaxy with the Gaia data. We have provided an indication of the power of the catalogue in analysing the correlations between age and kinematics throughout the Galaxy. Although our catalogue is a combination of multiple surveys, the selection function of the total catalogue can be found, as most of the constituent catalogues have well defined selection functions. A future contribution will model the published catalogue.

\section*{Acknowledgements}

JLS acknowledges the support of the Science and Technology Facilities Council (STFC).
PD would like to acknowledge support from the STFC (ST/N000919/1).
We thank Gyuchul Myeong for checking an early version of the catalogue, and Douglas Boubert and Eugene Vasiliev for useful conversations.
We acknowledge the use of Sergey Koposov's Whole Sky Database (WSDB), Andy Casey's Cannon code and Morgan Fouesneau's \texttt{ezpadova} tool. We thank Andy Casey for providing the correlations between spectroscopic parameters for the RAVE-On catalogue. This research made use of Astropy, a community-developed core Python package for Astronomy \citep{astropy} and packages in the \texttt{scipy} ecosystem \citep{scipy,matplotlib,ipython,pandas}.

This project was developed in part at the 2016 NYC Gaia Sprint, hosted by the Center for Computational Astrophysics at the Simons Foundation in New York City.

This work has made use of data from the European Space Agency (ESA) mission
{\it Gaia} (\url{https://www.cosmos.esa.int/web/gaia/iow_20180316}), processed by the {\it Gaia}
Data Processing and Analysis Consortium (DPAC,
\url{https://www.cosmos.esa.int/web/gaia/dpac/consortium}). Funding for the DPAC
has been provided by national institutions, in particular the institutions
participating in the {\it Gaia} Multilateral Agreement.

This publication makes use of data products from the Two Micron All Sky Survey, which is a joint project of the University of Massachusetts and the Infrared Processing and Analysis Center/California Institute of Technology, funded by the National Aeronautics and Space Administration and the National Science Foundation.

Funding for the Sloan Digital Sky Survey IV has been provided by the Alfred P. Sloan Foundation, the U.S. Department of Energy Office of Science, and the Participating Institutions. SDSS-IV acknowledges
support and resources from the Center for High-Performance Computing at
the University of Utah. The SDSS web site is www.sdss.org.

SDSS-IV is managed by the Astrophysical Research Consortium for the 
Participating Institutions of the SDSS Collaboration including the 
Brazilian Participation Group, the Carnegie Institution for Science, 
Carnegie Mellon University, the Chilean Participation Group, the French Participation Group, Harvard-Smithsonian Center for Astrophysics, 
Instituto de Astrof\'isica de Canarias, The Johns Hopkins University, 
Kavli Institute for the Physics and Mathematics of the Universe (IPMU) / 
University of Tokyo, Lawrence Berkeley National Laboratory, 
Leibniz Institut f\"ur Astrophysik Potsdam (AIP),  
Max-Planck-Institut f\"ur Astronomie (MPIA Heidelberg), 
Max-Planck-Institut f\"ur Astrophysik (MPA Garching), 
Max-Planck-Institut f\"ur Extraterrestrische Physik (MPE), 
National Astronomical Observatories of China, New Mexico State University, 
New York University, University of Notre Dame, 
Observat\'ario Nacional / MCTI, The Ohio State University, 
Pennsylvania State University, Shanghai Astronomical Observatory, 
United Kingdom Participation Group,
Universidad Nacional Aut\'onoma de M\'exico, University of Arizona, 
University of Colorado Boulder, University of Oxford, University of Portsmouth, 
University of Utah, University of Virginia, University of Washington, University of Wisconsin, 
Vanderbilt University, and Yale University.

Guoshoujing Telescope (the Large Sky Area Multi-Object Fiber Spectroscopic Telescope LAMOST) is a National Major Scientific Project built by the Chinese Academy of Sciences. Funding for the project has been provided by the National Development and Reform Commission. LAMOST is operated and managed by the National Astronomical Observatories, Chinese Academy of Sciences.

Funding for RAVE has been provided by: the Australian Astronomical Observatory; the Leibniz-Institut fuer Astrophysik Potsdam (AIP); the Australian National University; the Australian Research Council; the French National Research Agency; the German Research Foundation (SPP 1177 and SFB 881); the European Research Council (ERC-StG 240271 Galactica); the Istituto Nazionale di Astrofisica at Padova; The Johns Hopkins University; the National Science Foundation of the USA (AST-0908326); the W. M. Keck foundation; the Macquarie University; the Netherlands Research School for Astronomy; the Natural Sciences and Engineering Research Council of Canada; the Slovenian Research Agency; the Swiss National Science Foundation; the Science \& Technology Facilities Council of the UK; Opticon; Strasbourg Observatory; and the Universities of Groningen, Heidelberg and Sydney.
The RAVE web site is at https://www.rave-survey.org.




\bibliographystyle{mnras}
\bibliography{bibliography} 



\appendix

\section{Extinction coefficients}\label{Appendix::Extinction}
In our Bayesian distance pipeline, we require an extinction law to deredden any photometry used. This amounts to computing the set of coefficients $R(i)/R(V)$ for the photometric bands $i$. Given the $V$ band extinction $A_V$, the extinction in band $i$ is $A_i=R(i)/R(V)A_V$. The traditional definition of $R(V)$ is such that $A_V=R(V)E(B-V)$ where $E(B-V)$ is the selective extinction. As we work with the extinction maps of \cite{Green2017} who provide extinction in units of $E'$, we define $R(V)=A_V/E'$.

Here we choose to adopt the extinction curve $A(\lambda)$ from \cite{Schlafly2016} that was determined from APOGEE data. The curve was chosen to reproduce the observed extinction coefficients at the iso-extinction wavelengths for the Pan-STARRS and 2-MASS bands. \cite{Schlafly2016} parametrizes the extinction curve in terms of $x$ where $x=0$ corresponds approximately to $R_V=3.3$. \cite{Schlafly2016} finds that on average $R_V=3.3$ although there is variation of $\sim0.2$ across the APOGEE survey area. In the main body of the paper, we fix $R_V=3.3$ and hence choose $x=0$. The bluest band considered by \cite{Schlafly2016} was the Pan-STARRS $g_P$ band. Therefore, computing the extinction coefficients for bluer bands (particularly $u$) is an extrapolation.

Following \cite{Green2017}, we set the unknown grey component by insisting the extinction in the WISE $W_2$ band is zero. We set the scaling relative to the extinction $E'$ provided by \cite{Green2017} (chosen such that one unit of $E(g_P-r_P)$ for the Pan-STARRS bands produces one unit of \citeauthor{SFD} $E(B-V)$) by matching the coefficients for $g_P$ and $r_P$ provided in Table 1 of \cite{Green2017} at the iso-extinction wavelengths of \cite{Schlafly2016}. 
. We compute
\begin{equation}
R(i)=-\frac{2.5}{E'}\log_{10}\frac{\int\mathrm{d}\lambda\,
10^{-0.4E' A(\lambda)}\lambda S_i(\lambda)f(\lambda)}{\int\mathrm{d}\lambda \,\lambda S_i(\lambda)f(\lambda)},
\label{eqn::extinction}
\end{equation}
where $S_i(\lambda)$ is the response of bandpass $i$\footnote{Gaia DR2 photometric response curves downloaded from \href{www.cosmos.esa.int}{www.cosmos.esa.int}, Pan-STARRS from \href{http://ipp.ifa.hawaii.edu/ps1.filters/}{http://ipp.ifa.hawaii.edu/ps1.filters/}, 2MASS from \href{https://www.ipac.caltech.edu/2mass/releases/second/}{https://www.ipac.caltech.edu/2mass/releases/second/} (provides $\lambda S(\lambda)$), WISE from \href{http://wise2.ipac.caltech.edu/docs/release/prelim/expsup/}{http://wise2.ipac.caltech.edu/docs/release/prelim/expsup/}, Johnson $B$ and $V$ from \cite{MaizApellaniz2006} and SDSS from \href{http://classic.sdss.org/dr3/instruments/imager/filters/}{http://classic.sdss.org/dr3/instruments/imager/filters/}.} and $f(\lambda)$ the stellar model flux. From these coefficients and the selective extinction reported by \cite{Green2017}, the extinction in band $i$ is $A_i=R(i)E'$.

In Table~\ref{Table::Monochromatic} we provide the extinction coefficients evaluated using equation~\eqref{eqn::extinction} with $E'=0.01$ using for $f(\lambda)$ the \cite{CastelliKurucz2004} stellar model with solar metallicity, $\teff=4500\K$ and $\logg=4.5$. 

\begin{table}
\caption{Extinction coefficients from \protect\cite{Schlafly2016} extinction curve for use with \protect\cite{Green2017} extinction maps: $\lambda_\mathrm{eff}$ gives the effective wavelength, $R(\lambda_\mathrm{eff})$ the extinction coefficient evaluated at the effective wavelength and $R(\lambda)$ the extinction coefficient computed by integrating over a solar-metallicity stellar spectrum of effective temperature $4,500\K$ and surface gravity $\log g=4.5$. Given extinction in the $E'$ extinction units reported by \protect\cite{Green2017}, the extinction in band $i$ is $R(i)E'$.}
\input{extinction_coeff_2017_latex.dat}
\label{Table::Monochromatic}
\end{table}

\subsection{Variation of $R(G)$ with intrinsic colour}
As the Gaia $G$ band is broad, the coefficient provided in Table~\ref{Table::Monochromatic} is not appropriate for high extinction and for stars with intrinsic colours significantly different from that of a $\teff=4500\K$ star. To handle the latter of these issues, we evaluate equation~\ref{eqn::extinction} as a function of effective temperature and $R_V$ (we neglect any deviation with metallicity and surface gravity) using the solar metallicity and surface gravity $\logg=4.5$ stellar models of \cite{CastelliKurucz2004} (for $3500\K<\teff<50000\K$).
To the resulting $R(G)$, we fit polynomials of the form ($y^2$ and $y^3$ terms were unnecessary)
\begin{equation}
R(G)(x,y) = \bs{P}\cdot(1,x,y,x^2,x^2y,xy^2,xy,x^3).
\end{equation}
We use $x=\log_{10}\teff-4$ and $y=R_V-3.3$. Additionally, using a solar metallicity PARSEC isochrone \citep{Bressan2012,Chen2014,Tang2014,Chen2015} we can rewrite these polynomials with arguments $x=(J-K_s)$, $(g-r)$, $(g_P-r_P)$, $(B-V)$, $(G_\mathrm{BP}-G_\mathrm{RP})$, $(G-G_\mathrm{RP})$ and $(G_\mathrm{BP}-G)$.
The coefficients for this polynomial is given in Table~\ref{Table::RGPoly}. Throughout the main body of the paper we use $y=0$ so $R_V=3.3$.

\begin{table*}
\caption{$R(G)$ extinction coefficient polynomial coefficients where $R(G)(x,y) = \bs{P}\cdot(1,x,y,x^2,x^2y,xy^2,xy,x^3)$ where $A_G=R(G)E'$ for the extinction $E'$ reported by \protect\cite{Green2017} using \protect\cite{Schlafly2016} extinction curve.
Each row corresponds to a different $x$ given in the left column and $y=R_V-3.3$.
}
\begin{tabular}{lcccccccc}
\hline
 $\log_{10}(T_\mathrm{eff})-4$   & 2.9083 &  1.0268 & 0.0278 & -1.571  &  0.2401 & -0.0007 & -0.247  &  0.9734\\
 $(G-G_\mathrm{RP})$             & 2.9228 & -0.7571 & 0.022  & -0.164  &  0.008  &  0.0007 &  0.1633 &  0.081 \\
 $(G_\mathrm{BP}-G)$             & 2.9083 & -0.9529 & 0.0285 & -0.274  & -0.0392 &  0.0007 &  0.1937 &  0.3879\\
 $(G_\mathrm{BP}-G_\mathrm{RP})$ & 2.9176 & -0.4347 & 0.0255 & -0.0648 & -0.0041 &  0.0004 &  0.0893 &  0.0394\\
 $(J-K_s)$                       & 2.8808 & -1.0154 & 0.0302 &  0.2385 & -0.0696 &  0.0008 &  0.2245 & -0.0014\\
 $(g-r)$                         & 2.7549 & -0.5898 & 0.0525 &  0.3843 & -0.0425 &  0.0005 &  0.1456 & -0.2209\\
 $(g_\mathrm{P}-r_\mathrm{P})$   & 2.7433 & -0.6583 & 0.0552 &  0.4938 & -0.056  &  0.0006 &  0.165  & -0.3346\\
 $(B-V)$                         & 2.908  & -0.7492 & 0.0198 &  0.3956 & -0.0402 &  0.0005 &  0.1621 & -0.1553\\

\hline
\end{tabular}
\label{Table::RGPoly}
\end{table*}

\begin{table*}
\caption{$R(G)$ extinction coefficient polynomial coefficients where $R(G)(x,y) = \bs{P}\cdot(1,x,y,x^2,x^2y,y^2,xy^2,xy,x^3,y^3)$ where $A_G=R(G)E(B-V)_\mathrm{SFD}$ for the extinction $E(B-V)_\mathrm{SFD}$ reported by \protect\cite{SFD} using \protect\cite{Fitzpatrick1999} extinction curve and the correction of \protect\cite{Schlafly2011}.
Each row corresponds to a different $x$ given in the left column and $y=R_V-3.1$.
}
\begin{tabular}{lcccccccccc}
\hline
$\log_{10}(T_\mathrm{eff})-4$    &2.6241  &0.9392  &-0.1496  &-1.4703 &  0.1517  & 0.0525 &  0.0368  &-0.2465 &  0.9321 & -0.0092\\
$(G-G_\mathrm{RP})$              &2.6343  &-0.698  & -0.1641 & -0.1484 & -0.0126 & 0.0579 & -0.0356  & 0.1954 &  0.0846 & -0.0092\\
$(G_\mathrm{BP}-G)$              &2.6214  &-0.881  & -0.1601 & -0.2573 & -0.0424 & 0.0579 & -0.0365  & 0.2183 &  0.3736 & -0.0092\\
$(G_\mathrm{BP}-G_\mathrm{RP})$  &2.6298  &-0.4016 & -0.1619 & -0.0602 & -0.0071 & 0.0579 & -0.0182  & 0.1038 &  0.0382 & -0.0092\\
$(J-K_s)$                        &2.5951  &-0.9401 & -0.1587 &  0.226  & -0.0725 & 0.058  & -0.0401  & 0.2513 &  0.014  & -0.0092\\
$(g-r)$                          &2.4775  &-0.5453 & -0.1352 &  0.3646 & -0.0411 & 0.0549 & -0.0254  & 0.1604 & -0.2039 & -0.0092\\
$(g_\mathrm{P}-r_\mathrm{P})$    &2.4668  &-0.6083 & -0.1323 &  0.4688 & -0.0541 & 0.0545 & -0.0289  & 0.1818 & -0.3089 & -0.0092\\
$(B-V)$                          &2.6197  &-0.6955 & -0.1708 &  0.3719 & -0.0394 & 0.0602 & -0.0259  & 0.1774 & -0.1416 & -0.0092\\
\hline
\end{tabular}
\label{Table::RGPolySchlegel}
\end{table*}

\begin{figure}
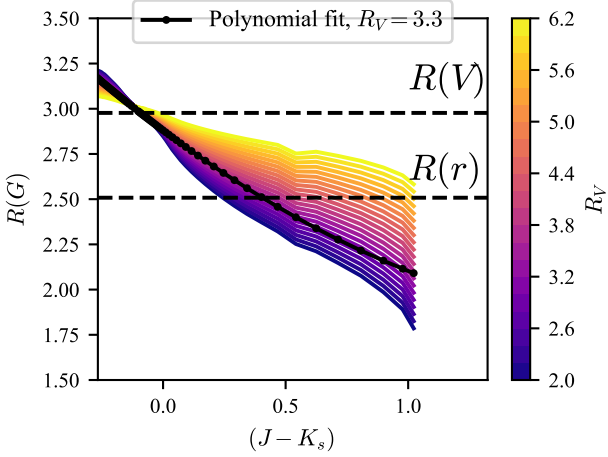

$$\includegraphics[width=\columnwidth]{{{figures/extinction_Gaia_JK}}}$$
\caption{Extinction coefficient for Gaia $G$-band: the extinction coefficient $R(G)$ is plotted as a function of $(J-K_s)$ coloured by the choice of \protect\cite{Schlafly2016} extinction law coefficient $R_V$. The black dots show the polynomial fit for $R_V=3.3$. The two horizontal dashed lines show the extinction coefficients for the narrow bands $r$ and $V$. This extinction coefficient is for use with the extinctions $E'$ reported by \protect\cite{Green2017} as $A_G=R(G)E'$.}
\label{Fig::ExtinctionGaia}
\end{figure}

The run of $R(G)$ with $(J-K_s)$ is shown in Fig.~\ref{Fig::ExtinctionGaia} coloured by choice of $R_V$. We also show the polynomial result for $R_V=3.3$ in black. We note that the variation of $R(G)$ with effective temperature is reasonably large ($\sim40\percent$) where for blue stars ($J-K_s=0$) $R(G)\approx R(V)=2.98$, for redder stars of $(J-K_s)=0.4$ $R(G)\approx R(r)=2.51$ and for very red stars ($J-K_s=1$) $R(G)$ is as low as $2.1$. 

As the Gaia $G$ band is broad, we also consider variation in $R(G)$ with the monochromatic extinction. For two models $\teff=(4500,10000)\K, \logg=4.5$, we compute equation~\eqref{eqn::extinction} for a range of $0.01<E'<1$. We fit a simple linear relation to $A_G/A_V$ against $A_V$ 
\begin{equation}
A_G/A_V = R(G)(1-c_G A_V),
\end{equation}
finding that for the $\teff=4500\K$ model $c_G=-0.031$ and for $\teff=10000\K$ model $c_G=-0.038$. We adopt the simple relation $c_G=-0.03$. Similarly we find for both $G_\mathrm{BP}$ and $G_\mathrm{RP}$, $c_{\mathrm{BP},\mathrm{RP}}=-0.01$. For all other bands these gradients are negligible.   

For completeness we provide $R(G)$ polynomial fits for use with \cite{SFD} extinction $E(B-V)_\mathrm{SFD}$ i.e. $A_i=R(i)E(B-V)_\mathrm{SFD}$. We utilize a \cite{Fitzpatrick1999} extinction law corrected for the \cite{SFD} units using \cite{Schlafly2011}. Again, we consider a range of stellar models and fit polynomials of the form
\begin{equation}
R(G)(x,y) = \bs{P}\cdot(1,x,y,x^2,x^2y,xy^2,xy,x^3),
\end{equation}
with $y=R_V-3.1$. The $G$ extinction is then computed using $A_G=R(G)E(B-V)_\mathrm{SFD}$.
The results are given in Table~\ref{Table::RGPolySchlegel}. For $G_\mathrm{RP}$ and $G_\mathrm{BP}$, we find coefficients $R(G_\mathrm{RP})=1.64$ and $R(G_\mathrm{BP})=2.73$.
Finally, we note that \cite{Babusiaux} also provides relations for $R(G)$, $R(G_\mathrm{BP})$ and $R(G_\mathrm{RP})$ in terms of the intrinsic colour and absolute extinction.
\section{Output catalogue format}\label{Sec::Catalogue}
In Table~\ref{Table::Catalogue} we detail the columns in the output catalogue available at \href{https://www.ast.cam.ac.uk/~jls/data/gaia_spectro.hdf5}{https://www.ast.cam.ac.uk/\~jls/data/gaia\_spectro.hdf5}. We give the column name, description and the unit of the quantity where applicable.
\begin{table*}
\caption{Format of provided catalogue. We give the names, description and unit of each provided quantity.}
\begin{tabularx}{0.8\textwidth}{lXc}
\hline
Name&Description&Unit\\
\hline
\texttt{survey}&String describing the spectroscopic survey from which the entry derives -- APOGEE, GALAH, GES, RAVEON, RAVEDR5, LAMOST, SEGUE&-\\
\texttt{raveid}&unique ID for the RAVE (and RAVE-ON) catalogue&-\\
\texttt{APOGEE\_ID}&unique ID for the APOGEE catalogue&-\\
\texttt{obsid}&unique ID for the LAMOST catalogue&-\\
\texttt{sobject\_id}&unique ID for the GALAH catalogue&-\\
\texttt{CNAME}&unique ID for the GES catalogue&-\\
\texttt{specobjid}&unique ID for the SEGUE catalogue&-\\
\texttt{source\_id}&Cross-matched Gaia DR2 source ID&-\\
\texttt{angular\_separation}&On-sky separation between spectroscopic catalogue entry and Gaia DR2 cross-match (epoch corrected if proper motion has been considered)&arcsec\\
\texttt{ra}&Right ascension from spectroscopic catalogue&degree\\
\texttt{dec}&Declination from spectroscopic catalogue&degree\\
\texttt{mag\_use}&Comma-separated string giving photometry used in pipeline: \texttt{J}, \texttt{H}, \texttt{K} are 2MASS bands, \texttt{G}, \texttt{GBP}, \texttt{GRP} are Gaia bands, \texttt{gP}, \texttt{rP}, \texttt{iP} are Pan-STARRS bands, \texttt{g}, \texttt{r}, \texttt{i} are SDSS bands, \texttt{Jv} ,\texttt{Hv}, \texttt{Kv} are VISTA bands&-\\
\hline
\textbf{Primary outputs}&&\\
\texttt{dm}&Distance modulus&mag\\
\texttt{dist}&Distance&kpc\\
\texttt{par}&Parallax (not Gaia DR2 parallax)&mas\\
\texttt{log10\_age}&Base-10 logarithm of age&$\log_{10}$(Gyr)\\
\texttt{mass}&(Initial) Mass&M$_\odot$\\
\texttt{Z}&Metallicity&dex\\
\texttt{log10\_av}&Base-10 logarithm of $V$-band extinction&$\log_{10}$(mag)\\
\texttt{log10\_teff}&Base-10 logarithm of effective temperature&$\log_{10}$(K)\\
\texttt{logg}&Surface gravity&$\log_{10}(\mathrm{cm}/\mathrm{s}^2)$\\
\hline
\textbf{Auxiliary outputs}&&\\
\texttt{l}&Galactic longitude&rad\\
\texttt{b}&Galactic latitude&rad\\
\texttt{s}&Distance derived from \texttt{dm} column&kpc\\
\texttt{vlos}&Line-of-sight velocity&km s$^{-1}$\\
\texttt{mu\_l}&Proper motion in Galactic longitude&mas yr$^{-1}$\\
\texttt{mu\_b}&Proper motion in Galactic latitude&mas yr$^{-1}$\\
\texttt{R}&Galactocentric cylindrical polar radius&kpc\\
\texttt{phi}&Galactocentric cylindrical polar angle (zero at solar position increasing in direction opposite to solar azimuthal velocity)&rad\\
\texttt{z}&Galactic height&kpc\\
\texttt{vR}&Galactocentric cylindrical radial velocity&km s$^{-1}$\\
\texttt{vphi}&Galactocentric azimuthal velocity (positive for Sun and decreasing \texttt{phi})&km s$^{-1}$\\
\texttt{vz}&Galactocentric vertical velocity&km s$^{-1}$\\
\texttt{JR}&Radial action&kpc km s$^{-1}$\\
\texttt{Lz}&$z$-component of angular momentum (positive for Sun)&kpc km s$^{-1}$\\
\texttt{Jz}&Vertical action&kpc km s$^{-1}$\\
\texttt{Rc}&Galactocentric radius of circular orbit with angular momentum \texttt{Lz}&kpc\\
\texttt{*\_err}&Corresponding uncertainty for each field&-\\
\texttt{*\_*\_corr}&Correlations in derived quantities&-\\
\hline
\textbf{Flags}&&\\
\texttt{flag}&non-zero if pipeline has failed (see text for meaning of each value)&-\\
\texttt{duplicated}&1 if duplicate Gaia \texttt{source\_id} (see text for how we select the source to keep)&-\\
\texttt{best}&1 if \texttt{flag}=0, \texttt{duplicated}=0 and valid Gaia \texttt{source\_id}&-\\
\hline
\end{tabularx}
\label{Table::Catalogue}
\end{table*}

\bsp	
\label{lastpage}
\end{document}

%% file: extinction_coeff_2017_latex.dat
\begin{tabular}{lcccc}\\Band&$\lambda_\mathrm{eff}/\,\angstrom$&$R(\lambda_\mathrm{eff})$&$R(\lambda)$&$A_\lambda/A_
V$\\
\hline 
$u$&$3607.7$&$4.781$&$4.869$&$1.636$\\
$B$&$4533.6$&$3.830$&$3.878$&$1.303$\\
$g$&$4811.9$&$3.576$&$3.613$&$1.214$\\
$g_\mathrm{P}$&$4958.8$&$3.449$&$3.481$&$1.169$\\
$G_\mathrm{BP}$&$5524.6$&$2.995$&$3.046$&$1.023$\\
$V$&$5560.0$&$2.969$&$2.977$&$1.000$\\
$r$&$6202.6$&$2.533$&$2.519$&$0.846$\\
$r_\mathrm{P}$&$6239.8$&$2.510$&$2.499$&$0.840$\\
$G$&$6791.6$&$2.201$&$2.294$&$0.771$\\
$i$&$7503.5$&$1.868$&$1.824$&$0.613$\\
$i_\mathrm{P}$&$7548.3$&$1.849$&$1.805$&$0.606$\\
$G_\mathrm{RP}$&$7782.1$&$1.752$&$1.737$&$0.584$\\
$G_\mathrm{RVS}$&$8597.2$&$1.449$&$1.393$&$0.468$\\
$z_\mathrm{P}$&$8679.3$&$1.421$&$1.368$&$0.460$\\
$z$&$8957.7$&$1.329$&$1.285$&$0.432$\\
$y_\mathrm{P}$&$9631.3$&$1.132$&$1.084$&$0.364$\\
$J$&$12365.1$&$0.662$&$0.632$&$0.212$\\
$H$&$16426.9$&$0.335$&$0.318$&$0.107$\\
$K_s$&$21526.0$&$0.170$&$0.158$&$0.053$\\
$W_1$&$33183.4$&$0.046$&$0.044$&$0.015$\\
$W_2$&$45550.9$&$-0.002$&$0.004$&$0.001$\\
\hline\end{tabular}

%% file: main.bbl
\begin{thebibliography}{}
\makeatletter
\relax
\def\mn@urlcharsother{\let\do\@makeother \do\$\do\&\do\#\do\^\do\_\do\%\do\~}
\def\mn@doi{\begingroup\mn@urlcharsother \@ifnextchar [ {\mn@doi@}
  {\mn@doi@[]}}
\def\mn@doi@[#1]#2{\def\@tempa{#1}\ifx\@tempa\@empty \href
  {http://dx.doi.org/#2} {doi:#2}\else \href {http://dx.doi.org/#2} {#1}\fi
  \endgroup}
\def\mn@eprint#1#2{\mn@eprint@#1:#2::\@nil}
\def\mn@eprint@arXiv#1{\href {http://arxiv.org/abs/#1} {{\tt arXiv:#1}}}
\def\mn@eprint@dblp#1{\href {http://dblp.uni-trier.de/rec/bibtex/#1.xml}
  {dblp:#1}}
\def\mn@eprint@#1:#2:#3:#4\@nil{\def\@tempa {#1}\def\@tempb {#2}\def\@tempc
  {#3}\ifx \@tempc \@empty \let \@tempc \@tempb \let \@tempb \@tempa \fi \ifx
  \@tempb \@empty \def\@tempb {arXiv}\fi \@ifundefined
  {mn@eprint@\@tempb}{\@tempb:\@tempc}{\expandafter \expandafter \csname
  mn@eprint@\@tempb\endcsname \expandafter{\@tempc}}}

\bibitem[\protect\citeauthoryear{{Abolfathi} et~al.,}{{Abolfathi}
  et~al.}{2018}]{SDSSDR14}
{Abolfathi} B.,  et~al., 2018, \mn@doi [\apjs] {10.3847/1538-4365/aa9e8a},
  \href {http://adsabs.harvard.edu/abs/2018ApJS..235...42A} {235, 42}

\bibitem[\protect\citeauthoryear{{Allende Prieto} et~al.,}{{Allende Prieto}
  et~al.}{2008}]{AllendePrieto2008}
{Allende Prieto} C.,  et~al., 2008, \mn@doi [\aj]
  {10.1088/0004-6256/136/5/2070}, \href
  {http://adsabs.harvard.edu/abs/2008AJ....136.2070A} {136, 2070}

\bibitem[\protect\citeauthoryear{{Andrae} et~al.,}{{Andrae}
  et~al.}{2018}]{Andrae2018}
{Andrae} R.,  et~al., 2018, \mn@doi [\aap] {10.1051/0004-6361/201732516}, \href
  {http://adsabs.harvard.edu/abs/2018A%26A...616A...8A} {616, A8}

\bibitem[\protect\citeauthoryear{{Anguiano} et~al.,}{{Anguiano}
  et~al.}{2018}]{Anguiano2018}
{Anguiano} B.,  et~al., 2018, preprint, \href
  {http://adsabs.harvard.edu/abs/2018arXiv180707625A} {} (\mn@eprint {arXiv}
  {1807.07625})

\bibitem[\protect\citeauthoryear{{Aumer} \& {Binney}}{{Aumer} \&
  {Binney}}{2009}]{AumerBinney}
{Aumer} M.,  {Binney} J.~J.,  2009, \mn@doi [\mnras]
  {10.1111/j.1365-2966.2009.15053.x}, \href
  {http://adsabs.harvard.edu/abs/2009MNRAS.397.1286A} {397, 1286}

\bibitem[\protect\citeauthoryear{{Aumer}, {Binney}  \& {Sch{\"o}nrich}}{{Aumer}
  et~al.}{2016a}]{AumerBinney2016a}
{Aumer} M.,  {Binney} J.,   {Sch{\"o}nrich} R.,  2016a, \mn@doi [\mnras]
  {10.1093/mnras/stw777}, \href
  {http://adsabs.harvard.edu/abs/2016MNRAS.459.3326A} {459, 3326}

\bibitem[\protect\citeauthoryear{{Aumer}, {Binney}  \& {Sch{\"o}nrich}}{{Aumer}
  et~al.}{2016b}]{AumerBinney2016b}
{Aumer} M.,  {Binney} J.,   {Sch{\"o}nrich} R.,  2016b, \mn@doi [\mnras]
  {10.1093/mnras/stw1639}, \href
  {http://adsabs.harvard.edu/abs/2016MNRAS.462.1697A} {462, 1697}

\bibitem[\protect\citeauthoryear{{Bailer-Jones}, {Rybizki}, {Fouesneau},
  {Mantelet}  \& {Andrae}}{{Bailer-Jones} et~al.}{2018}]{BailerJones2018}
{Bailer-Jones} C.~A.~L.,  {Rybizki} J.,  {Fouesneau} M.,  {Mantelet} G.,
  {Andrae} R.,  2018, \mn@doi [\aj] {10.3847/1538-3881/aacb21}, \href
  {http://adsabs.harvard.edu/abs/2018AJ....156...58B} {156, 58}

\bibitem[\protect\citeauthoryear{{Bensby}, {Feltzing}  \& {Oey}}{{Bensby}
  et~al.}{2014}]{Bensby2014}
{Bensby} T.,  {Feltzing} S.,   {Oey} M.~S.,  2014, \mn@doi [\aap]
  {10.1051/0004-6361/201322631}, \href
  {http://adsabs.harvard.edu/abs/2014A%26A...562A..71B} {562, A71}

\bibitem[\protect\citeauthoryear{Binney}{Binney}{2012}]{Binney2012}
Binney J.,  2012, \mn@doi [\mnras] {10.1111/j.1365-2966.2012.21757.x}, 426,
  1324

\bibitem[\protect\citeauthoryear{{Binney} et~al.,}{{Binney}
  et~al.}{2014}]{Binney2014}
{Binney} J.,  et~al., 2014, \mn@doi [\mnras] {10.1093/mnras/stt1896}, \href
  {http://adsabs.harvard.edu/abs/2014MNRAS.437..351B} {437, 351}

\bibitem[\protect\citeauthoryear{{Bland-Hawthorn} \&
  {Gerhard}}{{Bland-Hawthorn} \& {Gerhard}}{2016}]{BlandHawthornGerhard2016}
{Bland-Hawthorn} J.,  {Gerhard} O.,  2016, \mn@doi [\araa]
  {10.1146/annurev-astro-081915-023441}, \href
  {http://adsabs.harvard.edu/abs/2016ARA%26A..54..529B} {54, 529}

\bibitem[\protect\citeauthoryear{{Blanton} et~al.,}{{Blanton}
  et~al.}{2017}]{Blanton2017}
{Blanton} M.~R.,  et~al., 2017, \mn@doi [\aj] {10.3847/1538-3881/aa7567}, \href
  {http://adsabs.harvard.edu/abs/2017AJ....154...28B} {154, 28}

\bibitem[\protect\citeauthoryear{{Bovy}}{{Bovy}}{2017}]{Bovy2018}
{Bovy} J.,  2017, \mn@doi [\mnras] {10.1093/mnras/stx1277}, \href
  {http://adsabs.harvard.edu/abs/2017MNRAS.470.1360B} {470, 1360}

\bibitem[\protect\citeauthoryear{{Bovy}, {Rix}, {Green}, {Schlafly}  \&
  {Finkbeiner}}{{Bovy} et~al.}{2016}]{Bovy2016}
{Bovy} J.,  {Rix} H.-W.,  {Green} G.~M.,  {Schlafly} E.~F.,   {Finkbeiner}
  D.~P.,  2016, \mn@doi [\apj] {10.3847/0004-637X/818/2/130}, \href
  {http://adsabs.harvard.edu/abs/2016ApJ...818..130B} {818, 130}

\bibitem[\protect\citeauthoryear{{Bressan}, {Marigo}, {Girardi}, {Salasnich},
  {Dal Cero}, {Rubele}  \& {Nanni}}{{Bressan} et~al.}{2012}]{Bressan2012}
{Bressan} A.,  {Marigo} P.,  {Girardi} L.,  {Salasnich} B.,  {Dal Cero} C.,
  {Rubele} S.,   {Nanni} A.,  2012, \mn@doi [\mnras]
  {10.1111/j.1365-2966.2012.21948.x}, \href
  {http://adsabs.harvard.edu/abs/2012MNRAS.427..127B} {427, 127}

\bibitem[\protect\citeauthoryear{{Buder} et~al.,}{{Buder}
  et~al.}{2018}]{Buder2018}
{Buder} S.,  et~al., 2018, \mn@doi [\mnras] {10.1093/mnras/sty1281}, \href
  {http://adsabs.harvard.edu/abs/2018MNRAS.478.4513B} {478, 4513}

\bibitem[\protect\citeauthoryear{{Burnett} \& {Binney}}{{Burnett} \&
  {Binney}}{2010}]{BurnettBinney}
{Burnett} B.,  {Binney} J.,  2010, \mn@doi [\mnras]
  {10.1111/j.1365-2966.2010.16896.x}, \href
  {http://adsabs.harvard.edu/abs/2010MNRAS.407..339B} {407, 339}

\bibitem[\protect\citeauthoryear{{Casey}, {Hogg}, {Ness}, {Rix}, {Ho}  \&
  {Gilmore}}{{Casey} et~al.}{2016}]{Casey2016}
{Casey} A.~R.,  {Hogg} D.~W.,  {Ness} M.,  {Rix} H.-W.,  {Ho} A.~Q.,
  {Gilmore} G.,  2016, preprint, \href
  {http://adsabs.harvard.edu/abs/2016arXiv160303040C} {} (\mn@eprint {arXiv}
  {1603.03040})

\bibitem[\protect\citeauthoryear{{Casey} et~al.,}{{Casey}
  et~al.}{2017}]{Casey2017}
{Casey} A.~R.,  et~al., 2017, \mn@doi [\apj] {10.3847/1538-4357/aa69c2}, \href
  {http://adsabs.harvard.edu/abs/2017ApJ...840...59C} {840, 59}

\bibitem[\protect\citeauthoryear{{Castelli} \& {Kurucz}}{{Castelli} \&
  {Kurucz}}{2004}]{CastelliKurucz2004}
{Castelli} F.,  {Kurucz} R.~L.,  2004, ArXiv Astrophysics e-prints, \href
  {http://adsabs.harvard.edu/abs/2004astro.ph..5087C} {}

\bibitem[\protect\citeauthoryear{{Charbonnel}}{{Charbonnel}}{1994}]{Charbonnel1994}
{Charbonnel} C.,  1994, \aap, \href
  {http://adsabs.harvard.edu/abs/1994A%26A...282..811C} {282, 811}

\bibitem[\protect\citeauthoryear{{Chen}, {Girardi}, {Bressan}, {Marigo},
  {Barbieri}  \& {Kong}}{{Chen} et~al.}{2014}]{Chen2014}
{Chen} Y.,  {Girardi} L.,  {Bressan} A.,  {Marigo} P.,  {Barbieri} M.,   {Kong}
  X.,  2014, \mn@doi [\mnras] {10.1093/mnras/stu1605}, \href
  {http://adsabs.harvard.edu/abs/2014MNRAS.444.2525C} {444, 2525}

\bibitem[\protect\citeauthoryear{{Chen}, {Bressan}, {Girardi}, {Marigo}, {Kong}
   \& {Lanza}}{{Chen} et~al.}{2015}]{Chen2015}
{Chen} Y.,  {Bressan} A.,  {Girardi} L.,  {Marigo} P.,  {Kong} X.,   {Lanza}
  A.,  2015, \mn@doi [\mnras] {10.1093/mnras/stv1281}, \href
  {http://adsabs.harvard.edu/abs/2015MNRAS.452.1068C} {452, 1068}

\bibitem[\protect\citeauthoryear{{Cirasuolo} et~al.,}{{Cirasuolo}
  et~al.}{2012}]{MOONS}
{Cirasuolo} M.,  et~al., 2012, in Ground-based and Airborne Instrumentation for
  Astronomy IV. p. 84460S (\mn@eprint {arXiv} {1208.5780}),
  \mn@doi{10.1117/12.925871}

\bibitem[\protect\citeauthoryear{{Coronado}, {Rix}  \& {Trick}}{{Coronado}
  et~al.}{2018}]{Coronado2018}
{Coronado} J.,  {Rix} H.-W.,   {Trick} W.~H.,  2018, preprint, \href
  {http://adsabs.harvard.edu/abs/2018arXiv180407760C} {} (\mn@eprint {arXiv}
  {1804.07760})

\bibitem[\protect\citeauthoryear{{Cropper} et~al.,}{{Cropper}
  et~al.}{2018}]{GaiaDR2RVS1}
{Cropper} M.,  et~al., 2018, preprint, \href
  {http://adsabs.harvard.edu/abs/2018arXiv180409369C} {} (\mn@eprint {arXiv}
  {1804.09369})

\bibitem[\protect\citeauthoryear{Cui et~al.,}{Cui et~al.}{2012}]{Cui2012}
Cui X.-Q.,  et~al., 2012, Research in Astronomy and Astrophysics, 12, 1197

\bibitem[\protect\citeauthoryear{{Dalton} et~al.,}{{Dalton}
  et~al.}{2006}]{VISTA2}
{Dalton} G.~B.,  et~al., 2006, in Society of Photo-Optical Instrumentation
  Engineers (SPIE) Conference Series. p. 62690X, \mn@doi{10.1117/12.670018}

\bibitem[\protect\citeauthoryear{{Dalton} et~al.,}{{Dalton}
  et~al.}{2012}]{WEAVE}
{Dalton} G.,  et~al., 2012, in Ground-based and Airborne Instrumentation for
  Astronomy IV. p. 84460P, \mn@doi{10.1117/12.925950}

\bibitem[\protect\citeauthoryear{{Das} \& {Sanders}}{{Das} \&
  {Sanders}}{2018}]{DasSanders2018}
{Das} P.,  {Sanders} J.,  2018, preprint, \href
  {http://adsabs.harvard.edu/abs/2018arXiv180409596D} {} (\mn@eprint {arXiv}
  {1804.09596})

\bibitem[\protect\citeauthoryear{{De Silva} et~al.,}{{De Silva}
  et~al.}{2015}]{DeSilva2015}
{De Silva} G.~M.,  et~al., 2015, \mn@doi [\mnras] {10.1093/mnras/stv327}, \href
  {http://adsabs.harvard.edu/abs/2015MNRAS.449.2604D} {449, 2604}

\bibitem[\protect\citeauthoryear{{Deng} et~al.,}{{Deng}
  et~al.}{2012}]{Deng2012}
{Deng} L.-C.,  et~al., 2012, \mn@doi [Research in Astronomy and Astrophysics]
  {10.1088/1674-4527/12/7/003}, \href
  {http://adsabs.harvard.edu/abs/2012RAA....12..735D} {12, 735}

\bibitem[\protect\citeauthoryear{{Drimmel}, {Cabrera-Lavers}  \&
  {L{\'o}pez-Corredoira}}{{Drimmel} et~al.}{2003}]{Drimmel2003}
{Drimmel} R.,  {Cabrera-Lavers} A.,   {L{\'o}pez-Corredoira} M.,  2003, \mn@doi
  [\aap] {10.1051/0004-6361:20031070}, \href
  {http://adsabs.harvard.edu/abs/2003A%26A...409..205D} {409, 205}

\bibitem[\protect\citeauthoryear{{Emerson}, {McPherson}  \&
  {Sutherland}}{{Emerson} et~al.}{2006}]{VISTA1}
{Emerson} J.,  {McPherson} A.,   {Sutherland} W.,  2006, The Messenger, \href
  {http://adsabs.harvard.edu/abs/2006Msngr.126...41E} {126, 41}

\bibitem[\protect\citeauthoryear{{Evans} et~al.,}{{Evans}
  et~al.}{2018}]{GaiaDR2Photometry2}
{Evans} D.~W.,  et~al., 2018, \mn@doi [\aap] {10.1051/0004-6361/201832756},
  \href {http://adsabs.harvard.edu/abs/2018A%26A...616A...4E} {616, A4}

\bibitem[\protect\citeauthoryear{{Fitzpatrick}}{{Fitzpatrick}}{1999}]{Fitzpatrick1999}
{Fitzpatrick} E.~L.,  1999, \mn@doi [\pasp] {10.1086/316293}, \href
  {http://adsabs.harvard.edu/abs/1999PASP..111...63F} {111, 63}

\bibitem[\protect\citeauthoryear{{Freeman} \& {Bland-Hawthorn}}{{Freeman} \&
  {Bland-Hawthorn}}{2002}]{FreemanBlandHawthorn2002}
{Freeman} K.,  {Bland-Hawthorn} J.,  2002, \mn@doi [\araa]
  {10.1146/annurev.astro.40.060401.093840}, \href
  {http://adsabs.harvard.edu/abs/2002ARA%26A..40..487F} {40, 487}

\bibitem[\protect\citeauthoryear{{Gaia Collaboration} et~al.,}{{Gaia
  Collaboration} et~al.}{2016}]{Gaia2016}
{Gaia Collaboration} et~al., 2016, \mn@doi [\aap]
  {10.1051/0004-6361/201629272}, \href
  {http://adsabs.harvard.edu/abs/2016A%26A...595A...1G} {595, A1}

\bibitem[\protect\citeauthoryear{{Gaia Collaboration} et~al.,}{{Gaia
  Collaboration} et~al.}{2018a}]{GaiaDR2}
{Gaia Collaboration} et~al., 2018a, \mn@doi [\aap]
  {10.1051/0004-6361/201833051}, \href
  {http://adsabs.harvard.edu/abs/2018A%26A...616A...1G} {616, A1}

\bibitem[\protect\citeauthoryear{{Gaia Collaboration} et~al.,}{{Gaia
  Collaboration} et~al.}{2018b}]{Babusiaux}
{Gaia Collaboration} et~al., 2018b, \mn@doi [\aap]
  {10.1051/0004-6361/201832843}, \href
  {http://adsabs.harvard.edu/abs/2018A%26A...616A..10G} {616, A10}

\bibitem[\protect\citeauthoryear{{Gaia Collaboration} et~al.,}{{Gaia
  Collaboration} et~al.}{2018c}]{GaiaMWKinematics}
{Gaia Collaboration} et~al., 2018c, \mn@doi [\aap]
  {10.1051/0004-6361/201832865}, \href
  {http://adsabs.harvard.edu/abs/2018A%26A...616A..11G} {616, A11}

\bibitem[\protect\citeauthoryear{{Gaia Collaboration} et~al.,}{{Gaia
  Collaboration} et~al.}{2018d}]{Helmi2018}
{Gaia Collaboration} et~al., 2018d, \mn@doi [\aap]
  {10.1051/0004-6361/201832698}, \href
  {http://adsabs.harvard.edu/abs/2018A%26A...616A..12G} {616, A12}

\bibitem[\protect\citeauthoryear{{Garc{\'{\i}}a P{\'e}rez}
  et~al.,}{{Garc{\'{\i}}a P{\'e}rez} et~al.}{2016}]{GarciaPerez2016}
{Garc{\'{\i}}a P{\'e}rez} A.~E.,  et~al., 2016, \mn@doi [\aj]
  {10.3847/0004-6256/151/6/144}, \href
  {http://adsabs.harvard.edu/abs/2016AJ....151..144G} {151, 144}

\bibitem[\protect\citeauthoryear{{Gilmore} et~al.,}{{Gilmore}
  et~al.}{2012}]{Gilmore2012}
{Gilmore} G.,  et~al., 2012, The Messenger, \href
  {http://adsabs.harvard.edu/abs/2012Msngr.147...25G} {147, 25}

\bibitem[\protect\citeauthoryear{{Green} et~al.,}{{Green}
  et~al.}{2018}]{Green2017}
{Green} G.~M.,  et~al., 2018, \mn@doi [\mnras] {10.1093/mnras/sty1008}, \href
  {http://adsabs.harvard.edu/abs/2018MNRAS.478..651G} {478, 651}

\bibitem[\protect\citeauthoryear{{Gunn} et~al.,}{{Gunn}
  et~al.}{2006}]{Gunn2006}
{Gunn} J.~E.,  et~al., 2006, \mn@doi [\aj] {10.1086/500975}, \href
  {http://adsabs.harvard.edu/abs/2006AJ....131.2332G} {131, 2332}

\bibitem[\protect\citeauthoryear{{Ho} et~al.,}{{Ho} et~al.}{2017}]{Ho2017}
{Ho} A.~Y.~Q.,  et~al., 2017, \mn@doi [\apj] {10.3847/1538-4357/836/1/5}, \href
  {http://adsabs.harvard.edu/abs/2017ApJ...836....5H} {836, 5}

\bibitem[\protect\citeauthoryear{{Hoffman} \& {Gelman}}{{Hoffman} \&
  {Gelman}}{2011}]{NUTS}
{Hoffman} M.~D.,  {Gelman} A.,  2011, preprint, \href
  {http://adsabs.harvard.edu/abs/2011arXiv1111.4246H} {} (\mn@eprint {arXiv}
  {1111.4246})

\bibitem[\protect\citeauthoryear{{Holmberg}, {Nordstr{\"o}m}  \&
  {Andersen}}{{Holmberg} et~al.}{2009}]{Holmberg2009}
{Holmberg} J.,  {Nordstr{\"o}m} B.,   {Andersen} J.,  2009, \mn@doi [\aap]
  {10.1051/0004-6361/200811191}, \href
  {http://adsabs.harvard.edu/abs/2009A%26A...501..941H} {501, 941}

\bibitem[\protect\citeauthoryear{{Howes}, {Lindegren}, {Feltzing}, {Church}  \&
  {Bensby}}{{Howes} et~al.}{2018}]{Howes2018}
{Howes} L.~M.,  {Lindegren} L.,  {Feltzing} S.,  {Church} R.~P.,   {Bensby} T.,
   2018, preprint, \href {http://adsabs.harvard.edu/abs/2018arXiv180408321H} {}
  (\mn@eprint {arXiv} {1804.08321})

\bibitem[\protect\citeauthoryear{Hunter}{Hunter}{2007}]{matplotlib}
Hunter J.~D.,  2007, \mn@doi [Computing In Science \& Engineering]
  {10.1109/MCSE.2007.55}, 9, 90

\bibitem[\protect\citeauthoryear{Jones, Oliphant, Peterson  et~al.}{Jones
  et~al.}{2001}]{scipy}
Jones E.,  Oliphant T.,  Peterson P.,   et~al., 2001, {SciPy}: Open source
  scientific tools for {Python}, \url {http://www.scipy.org/}

\bibitem[\protect\citeauthoryear{{J{\o}rgensen} \& {Lindegren}}{{J{\o}rgensen}
  \& {Lindegren}}{2005}]{JorgensenLindegren2005}
{J{\o}rgensen} B.~R.,  {Lindegren} L.,  2005, \mn@doi [\aap]
  {10.1051/0004-6361:20042185}, \href
  {http://adsabs.harvard.edu/abs/2005A%26A...436..127J} {436, 127}

\bibitem[\protect\citeauthoryear{{Kollmeier} et~al.,}{{Kollmeier}
  et~al.}{2017}]{SDSSV}
{Kollmeier} J.~A.,  et~al., 2017, preprint, \href
  {http://adsabs.harvard.edu/abs/2017arXiv171103234K} {} (\mn@eprint {arXiv}
  {1711.03234})

\bibitem[\protect\citeauthoryear{{Kordopatis} et~al.,}{{Kordopatis}
  et~al.}{2013}]{Kordopatis}
{Kordopatis} G.,  et~al., 2013, \mn@doi [\aj] {10.1088/0004-6256/146/5/134},
  \href {http://adsabs.harvard.edu/abs/2013AJ....146..134K} {146, 134}

\bibitem[\protect\citeauthoryear{{Kroupa}, {Tout}  \& {Gilmore}}{{Kroupa}
  et~al.}{1993}]{Kroupa}
{Kroupa} P.,  {Tout} C.~A.,   {Gilmore} G.,  1993, \mn@doi [\mnras]
  {10.1093/mnras/262.3.545}, \href
  {http://adsabs.harvard.edu/abs/1993MNRAS.262..545K} {262, 545}

\bibitem[\protect\citeauthoryear{{Kunder} et~al.,}{{Kunder}
  et~al.}{2017}]{Kunder2017}
{Kunder} A.,  et~al., 2017, \mn@doi [\aj] {10.3847/1538-3881/153/2/75}, \href
  {http://adsabs.harvard.edu/abs/2017AJ....153...75K} {153, 75}

\bibitem[\protect\citeauthoryear{{Lee} et~al.,}{{Lee} et~al.}{2008a}]{Lee2008a}
{Lee} Y.~S.,  et~al., 2008a, \mn@doi [\aj] {10.1088/0004-6256/136/5/2022},
  \href {http://adsabs.harvard.edu/abs/2008AJ....136.2022L} {136, 2022}

\bibitem[\protect\citeauthoryear{{Lee} et~al.,}{{Lee} et~al.}{2008b}]{Lee2008b}
{Lee} Y.~S.,  et~al., 2008b, \mn@doi [\aj] {10.1088/0004-6256/136/5/2050},
  \href {http://adsabs.harvard.edu/abs/2008AJ....136.2050L} {136, 2050}

\bibitem[\protect\citeauthoryear{{Lindegren} et~al.,}{{Lindegren}
  et~al.}{2018}]{GaiaDR2Astrometry}
{Lindegren} L.,  et~al., 2018, \mn@doi [\aap] {10.1051/0004-6361/201832727},
  \href {http://adsabs.harvard.edu/abs/2018A%26A...616A...2L} {616, A2}

\bibitem[\protect\citeauthoryear{{Luri} et~al.,}{{Luri}
  et~al.}{2018}]{Luri2018}
{Luri} X.,  et~al., 2018, \mn@doi [\aap] {10.1051/0004-6361/201832964}, \href
  {http://adsabs.harvard.edu/abs/2018A%26A...616A...9L} {616, A9}

\bibitem[\protect\citeauthoryear{{Mackereth} et~al.,}{{Mackereth}
  et~al.}{2017}]{Mackereth2017}
{Mackereth} J.~T.,  et~al., 2017, \mn@doi [\mnras] {10.1093/mnras/stx1774},
  \href {http://adsabs.harvard.edu/abs/2017MNRAS.471.3057M} {471, 3057}

\bibitem[\protect\citeauthoryear{{Magnier} et~al.,}{{Magnier}
  et~al.}{2013}]{Magnier2013}
{Magnier} E.~A.,  et~al., 2013, \mn@doi [\apjs] {10.1088/0067-0049/205/2/20},
  \href {http://adsabs.harvard.edu/abs/2013ApJS..205...20M} {205, 20}

\bibitem[\protect\citeauthoryear{{Ma{\'\i}z Apell{\'a}niz}}{{Ma{\'\i}z
  Apell{\'a}niz}}{2006}]{MaizApellaniz2006}
{Ma{\'\i}z Apell{\'a}niz} J.,  2006, \mn@doi [\aj] {10.1086/499158}, \href
  {https://ui.adsabs.harvard.edu/#abs/2006AJ....131.1184M} {131, 1184}

\bibitem[\protect\citeauthoryear{{Majewski} et~al.,}{{Majewski}
  et~al.}{2017}]{APOGEE}
{Majewski} S.~R.,  et~al., 2017, \mn@doi [\aj] {10.3847/1538-3881/aa784d},
  \href {http://adsabs.harvard.edu/abs/2017AJ....154...94M} {154, 94}

\bibitem[\protect\citeauthoryear{{Mar{\'{\i}}n-Franch}
  et~al.,}{{Mar{\'{\i}}n-Franch} et~al.}{2009}]{MarinFranch2009}
{Mar{\'{\i}}n-Franch} A.,  et~al., 2009, \mn@doi [\apj]
  {10.1088/0004-637X/694/2/1498}, \href
  {http://adsabs.harvard.edu/abs/2009ApJ...694.1498M} {694, 1498}

\bibitem[\protect\citeauthoryear{{Marshall}, {Robin}, {Reyl{\'e}}, {Schultheis}
   \& {Picaud}}{{Marshall} et~al.}{2006}]{Marshall2006}
{Marshall} D.~J.,  {Robin} A.~C.,  {Reyl{\'e}} C.,  {Schultheis} M.,   {Picaud}
  S.,  2006, \mn@doi [\aap] {10.1051/0004-6361:20053842}, \href
  {http://adsabs.harvard.edu/abs/2006A%26A...453..635M} {453, 635}

\bibitem[\protect\citeauthoryear{{Martell} et~al.,}{{Martell}
  et~al.}{2017}]{Martell2017}
{Martell} S.~L.,  et~al., 2017, \mn@doi [\mnras] {10.1093/mnras/stw2835}, \href
  {http://adsabs.harvard.edu/abs/2017MNRAS.465.3203M} {465, 3203}

\bibitem[\protect\citeauthoryear{{Martig}, {Minchev}  \& {Flynn}}{{Martig}
  et~al.}{2014}]{Martig2014}
{Martig} M.,  {Minchev} I.,   {Flynn} C.,  2014, \mn@doi [\mnras]
  {10.1093/mnras/stu1322}, \href
  {http://adsabs.harvard.edu/abs/2014MNRAS.443.2452M} {443, 2452}

\bibitem[\protect\citeauthoryear{{Martig} et~al.,}{{Martig}
  et~al.}{2016}]{Martig2016}
{Martig} M.,  et~al., 2016, \mn@doi [\mnras] {10.1093/mnras/stv2830}, \href
  {http://adsabs.harvard.edu/abs/2016MNRAS.456.3655M} {456, 3655}

\bibitem[\protect\citeauthoryear{{Masseron} \& {Gilmore}}{{Masseron} \&
  {Gilmore}}{2015}]{Masseron2015}
{Masseron} T.,  {Gilmore} G.,  2015, \mn@doi [\mnras] {10.1093/mnras/stv1731},
  \href {http://adsabs.harvard.edu/abs/2015MNRAS.453.1855M} {453, 1855}

\bibitem[\protect\citeauthoryear{McKinney}{McKinney}{2010}]{pandas}
McKinney W.,  2010, in van~der Walt S.,  Millman J.,  eds, Proceedings of the
  9th Python in Science Conference. pp 51 -- 56

\bibitem[\protect\citeauthoryear{{McMillan} et~al.,}{{McMillan}
  et~al.}{2018}]{McMillan2017}
{McMillan} P.~J.,  et~al., 2018, \mn@doi [\mnras] {10.1093/mnras/sty990}, \href
  {http://adsabs.harvard.edu/abs/2018MNRAS.477.5279M} {477, 5279}

\bibitem[\protect\citeauthoryear{{Minchev}, {Martig}, {Streich}, {Scannapieco},
  {de Jong}  \& {Steinmetz}}{{Minchev} et~al.}{2015}]{Minchev2015}
{Minchev} I.,  {Martig} M.,  {Streich} D.,  {Scannapieco} C.,  {de Jong} R.~S.,
    {Steinmetz} M.,  2015, \mn@doi [\apjl] {10.1088/2041-8205/804/1/L9}, \href
  {http://adsabs.harvard.edu/abs/2015ApJ...804L...9M} {804, L9}

\bibitem[\protect\citeauthoryear{{Mints} \& {Hekker}}{{Mints} \&
  {Hekker}}{2018}]{Mints2018}
{Mints} A.,  {Hekker} S.,  2018, preprint, \href
  {http://adsabs.harvard.edu/abs/2018arXiv180406578M} {} (\mn@eprint {arXiv}
  {1804.06578})

\bibitem[\protect\citeauthoryear{{Ness}, {Hogg}, {Rix}, {Martig},
  {Pinsonneault}  \& {Ho}}{{Ness} et~al.}{2016}]{Ness2016}
{Ness} M.,  {Hogg} D.~W.,  {Rix} H.-W.,  {Martig} M.,  {Pinsonneault} M.~H.,
  {Ho} A.~Y.~Q.,  2016, \mn@doi [\apj] {10.3847/0004-637X/823/2/114}, \href
  {http://adsabs.harvard.edu/abs/2016ApJ...823..114N} {823, 114}

\bibitem[\protect\citeauthoryear{P\'erez \& Granger}{P\'erez \&
  Granger}{2007}]{ipython}
P\'erez F.,  Granger B.~E.,  2007, \mn@doi [Computing in Science and
  Engineering] {10.1109/MCSE.2007.53}, 9, 21

\bibitem[\protect\citeauthoryear{{Pinsonneault} et~al.,}{{Pinsonneault}
  et~al.}{2014}]{Pinsonneault2014}
{Pinsonneault} M.~H.,  et~al., 2014, \mn@doi [\apjs]
  {10.1088/0067-0049/215/2/19}, \href
  {http://adsabs.harvard.edu/abs/2014ApJS..215...19P} {215, 19}

\bibitem[\protect\citeauthoryear{{Pont} \& {Eyer}}{{Pont} \&
  {Eyer}}{2004}]{PontEyer2004}
{Pont} F.,  {Eyer} L.,  2004, \mn@doi [\mnras]
  {10.1111/j.1365-2966.2004.07780.x}, \href
  {http://adsabs.harvard.edu/abs/2004MNRAS.351..487P} {351, 487}

\bibitem[\protect\citeauthoryear{{Queiroz} et~al.,}{{Queiroz}
  et~al.}{2018}]{Queiroz2018}
{Queiroz} A.~B.~A.,  et~al., 2018, \mn@doi [\mnras] {10.1093/mnras/sty330},
  \href {http://adsabs.harvard.edu/abs/2018MNRAS.476.2556Q} {476, 2556}

\bibitem[\protect\citeauthoryear{{Riello} et~al.,}{{Riello}
  et~al.}{2018}]{GaiaDR2Photometry1}
{Riello} M.,  et~al., 2018, \mn@doi [\aap] {10.1051/0004-6361/201832712}, \href
  {http://adsabs.harvard.edu/abs/2018A%26A...616A...3R} {616, A3}

\bibitem[\protect\citeauthoryear{{Riess} et~al.,}{{Riess}
  et~al.}{2018}]{Riess2018}
{Riess} A.~G.,  et~al., 2018, \mn@doi [\apj] {10.3847/1538-4357/aac82e}, \href
  {http://adsabs.harvard.edu/abs/2018ApJ...861..126R} {861, 126}

\bibitem[\protect\citeauthoryear{{Robin}, {Marshall}, {Schultheis}  \&
  {Reyl{\'e}}}{{Robin} et~al.}{2012}]{Robin2012}
{Robin} A.~C.,  {Marshall} D.~J.,  {Schultheis} M.,   {Reyl{\'e}} C.,  2012,
  \mn@doi [\aap] {10.1051/0004-6361/201116512}, \href
  {http://adsabs.harvard.edu/abs/2012A%26A...538A.106R} {538, A106}

\bibitem[\protect\citeauthoryear{{Salaris}, {Chieffi}  \&
  {Straniero}}{{Salaris} et~al.}{1993}]{Salaris1993}
{Salaris} M.,  {Chieffi} A.,   {Straniero} O.,  1993, \mn@doi [\apj]
  {10.1086/173105}, \href {http://adsabs.harvard.edu/abs/1993ApJ...414..580S}
  {414, 580}

\bibitem[\protect\citeauthoryear{Salvatier, Wiecki  \& Fonnesbeck}{Salvatier
  et~al.}{2016}]{Salvatier2016}
Salvatier J.,  Wiecki T.~V.,   Fonnesbeck C.,  2016, \mn@doi [{PeerJ} Computer
  Science] {10.7717/peerj-cs.55}, 2, e55

\bibitem[\protect\citeauthoryear{Sanders \& Binney}{Sanders \&
  Binney}{2016}]{SandersBinney2016}
Sanders J.,  Binney J.,  2016, \mn@doi [\mnras] {10.1093/mnras/stw106}, 457,
  2107

\bibitem[\protect\citeauthoryear{{Sartoretti} et~al.,}{{Sartoretti}
  et~al.}{2018}]{GaiaDR2RVS2}
{Sartoretti} P.,  et~al., 2018, \mn@doi [\aap] {10.1051/0004-6361/201832836},
  \href {http://adsabs.harvard.edu/abs/2018A%26A...616A...6S} {616, A6}

\bibitem[\protect\citeauthoryear{{Schlafly} \& {Finkbeiner}}{{Schlafly} \&
  {Finkbeiner}}{2011}]{Schlafly2011}
{Schlafly} E.~F.,  {Finkbeiner} D.~P.,  2011, \mn@doi [\apj]
  {10.1088/0004-637X/737/2/103}, \href
  {http://adsabs.harvard.edu/abs/2011ApJ...737..103S} {737, 103}

\bibitem[\protect\citeauthoryear{{Schlafly} et~al.,}{{Schlafly}
  et~al.}{2016}]{Schlafly2016}
{Schlafly} E.~F.,  et~al., 2016, \mn@doi [\apj] {10.3847/0004-637X/821/2/78},
  \href {http://adsabs.harvard.edu/abs/2016ApJ...821...78S} {821, 78}

\bibitem[\protect\citeauthoryear{{Schlegel}, {Finkbeiner}  \&
  {Davis}}{{Schlegel} et~al.}{1998}]{SFD}
{Schlegel} D.~J.,  {Finkbeiner} D.~P.,   {Davis} M.,  1998, \mn@doi [\apj]
  {10.1086/305772}, \href {http://adsabs.harvard.edu/abs/1998ApJ...500..525S}
  {500, 525}

\bibitem[\protect\citeauthoryear{{Sch{\"o}nrich}, {Binney}  \&
  {Dehnen}}{{Sch{\"o}nrich} et~al.}{2010}]{Schoenrich2010}
{Sch{\"o}nrich} R.,  {Binney} J.,   {Dehnen} W.,  2010, \mn@doi [\mnras]
  {10.1111/j.1365-2966.2010.16253.x}, \href
  {http://adsabs.harvard.edu/abs/2010MNRAS.403.1829S} {403, 1829}

\bibitem[\protect\citeauthoryear{Sharma, Bland-Hawthorn, Johnston  \&
  Binney}{Sharma et~al.}{2011}]{Sharma2011}
Sharma S.,  Bland-Hawthorn J.,  Johnston K.,   Binney J.,  2011, \mn@doi [\apj]
  {10.1088/0004-637X/730/1/3}, 730, 3

\bibitem[\protect\citeauthoryear{{Simion}, {Belokurov}, {Irwin}, {Koposov},
  {Gonzalez-Fernandez}, {Robin}, {Shen}  \& {Li}}{{Simion}
  et~al.}{2017}]{Simion2017}
{Simion} I.~T.,  {Belokurov} V.,  {Irwin} M.,  {Koposov} S.~E.,
  {Gonzalez-Fernandez} C.,  {Robin} A.~C.,  {Shen} J.,   {Li} Z.-Y.,  2017,
  \mn@doi [\mnras] {10.1093/mnras/stx1832}, \href
  {http://adsabs.harvard.edu/abs/2017MNRAS.471.4323S} {471, 4323}

\bibitem[\protect\citeauthoryear{{Skrutskie} et~al.,}{{Skrutskie}
  et~al.}{2006}]{2MASS}
{Skrutskie} M.~F.,  et~al., 2006, \mn@doi [\aj] {10.1086/498708}, \href
  {http://adsabs.harvard.edu/abs/2006AJ....131.1163S} {131, 1163}

\bibitem[\protect\citeauthoryear{{Smolinski} et~al.,}{{Smolinski}
  et~al.}{2011}]{Smolinski2011}
{Smolinski} J.~P.,  et~al., 2011, \mn@doi [\aj] {10.1088/0004-6256/141/3/89},
  \href {http://adsabs.harvard.edu/abs/2011AJ....141...89S} {141, 89}

\bibitem[\protect\citeauthoryear{{Soderblom}}{{Soderblom}}{2010}]{Soderblom2010}
{Soderblom} D.~R.,  2010, \mn@doi [\araa]
  {10.1146/annurev-astro-081309-130806}, \href
  {http://adsabs.harvard.edu/abs/2010ARA%26A..48..581S} {48, 581}

\bibitem[\protect\citeauthoryear{{Steinmetz} et~al.,}{{Steinmetz}
  et~al.}{2006}]{Steinmetz2006}
{Steinmetz} M.,  et~al., 2006, \mn@doi [\aj] {10.1086/506564}, \href
  {http://adsabs.harvard.edu/abs/2006AJ....132.1645S} {132, 1645}

\bibitem[\protect\citeauthoryear{{Tang}, {Bressan}, {Rosenfield}, {Slemer},
  {Marigo}, {Girardi}  \& {Bianchi}}{{Tang} et~al.}{2014}]{Tang2014}
{Tang} J.,  {Bressan} A.,  {Rosenfield} P.,  {Slemer} A.,  {Marigo} P.,
  {Girardi} L.,   {Bianchi} L.,  2014, \mn@doi [\mnras]
  {10.1093/mnras/stu2029}, \href
  {http://adsabs.harvard.edu/abs/2014MNRAS.445.4287T} {445, 4287}

\bibitem[\protect\citeauthoryear{{Tayar} et~al.,}{{Tayar}
  et~al.}{2015}]{Tayar2015}
{Tayar} J.,  et~al., 2015, \mn@doi [\apj] {10.1088/0004-637X/807/1/82}, \href
  {http://adsabs.harvard.edu/abs/2015ApJ...807...82T} {807, 82}

\bibitem[\protect\citeauthoryear{{The Astropy Collaboration} et~al.,}{{The
  Astropy Collaboration} et~al.}{2018}]{astropy}
{The Astropy Collaboration} et~al., 2018, \mn@doi [\aj]
  {10.3847/1538-3881/aabc4f}, \href
  {http://adsabs.harvard.edu/abs/2018AJ....156..123T} {156, 123}

\bibitem[\protect\citeauthoryear{{Vrard}, {Mosser}  \& {Samadi}}{{Vrard}
  et~al.}{2016}]{Vrard2016}
{Vrard} M.,  {Mosser} B.,   {Samadi} R.,  2016, \mn@doi [\aap]
  {10.1051/0004-6361/201527259}, \href
  {http://adsabs.harvard.edu/abs/2016A%26A...588A..87V} {588, A87}

\bibitem[\protect\citeauthoryear{{Wilson} et~al.,}{{Wilson}
  et~al.}{2010}]{APOGEEspectrograph}
{Wilson} J.~C.,  et~al., 2010, in Ground-based and Airborne Instrumentation for
  Astronomy III. p. 77351C, \mn@doi{10.1117/12.856708}

\bibitem[\protect\citeauthoryear{{Wu} et~al.,}{{Wu} et~al.}{2018}]{Wu2018}
{Wu} Y.,  et~al., 2018, \mn@doi [\mnras] {10.1093/mnras/stx3296}, \href
  {http://adsabs.harvard.edu/abs/2018MNRAS.475.3633W} {475, 3633}

\bibitem[\protect\citeauthoryear{{Xiang} et~al.,}{{Xiang}
  et~al.}{2015}]{Xiang2015}
{Xiang} M.-S.,  et~al., 2015, \mn@doi [Research in Astronomy and Astrophysics]
  {10.1088/1674-4527/15/8/009}, \href
  {http://adsabs.harvard.edu/abs/2015RAA....15.1209X} {15, 1209}

\bibitem[\protect\citeauthoryear{{Xiang} et~al.,}{{Xiang}
  et~al.}{2017}]{Xiang2017}
{Xiang} M.,  et~al., 2017, \mn@doi [\apjs] {10.3847/1538-4365/aa80e4}, \href
  {http://adsabs.harvard.edu/abs/2017ApJS..232....2X} {232, 2}

\bibitem[\protect\citeauthoryear{{Yanny} et~al.,}{{Yanny}
  et~al.}{2009}]{Yanny2009}
{Yanny} B.,  et~al., 2009, \mn@doi [\aj] {10.1088/0004-6256/137/5/4377}, \href
  {http://adsabs.harvard.edu/abs/2009AJ....137.4377Y} {137, 4377}

\bibitem[\protect\citeauthoryear{{Zhao}, {Zhao}, {Chu}, {Jing}  \&
  {Deng}}{{Zhao} et~al.}{2012}]{Zhao2012}
{Zhao} G.,  {Zhao} Y.,  {Chu} Y.,  {Jing} Y.,   {Deng} L.,  2012, preprint,
  \href {http://adsabs.harvard.edu/abs/2012arXiv1206.3569Z} {} (\mn@eprint
  {arXiv} {1206.3569})

\bibitem[\protect\citeauthoryear{{da Silva} et~al.,}{{da Silva}
  et~al.}{2006}]{deSilva2006}
{da Silva} L.,  et~al., 2006, \mn@doi [\aap] {10.1051/0004-6361:20065105},
  \href {http://adsabs.harvard.edu/abs/2006A%26A...458..609D} {458, 609}

\bibitem[\protect\citeauthoryear{{de Jong} et~al.,}{{de Jong}
  et~al.}{2016}]{4MOST}
{de Jong} R.~S.,  et~al., 2016, in Ground-based and Airborne Instrumentation
  for Astronomy VI. p. 99081O, \mn@doi{10.1117/12.2232832}

\makeatother
\end{thebibliography}
